\documentclass[3p,preprint,amsmath,amssymb]{revtex4-1}
\usepackage[utf8]{inputenc}
\usepackage{amsmath}
\usepackage{graphicx}
\usepackage{esint}

\usepackage{float}
\usepackage{dcolumn}
\usepackage{caption}
\usepackage{bm}
\usepackage{color}
\usepackage{upgreek}
\usepackage{braket}
\usepackage{ulem}
\usepackage{url}
\usepackage{hyperref}

\usepackage{setspace}

\usepackage{tabularx,booktabs}

\newcolumntype{Y}{>{\centering\arraybackslash}X}

\newcommand{\cm}{cm$^{-1}$}
\newcommand{\AnO}{A\&O}

\newcolumntype{L}[1]{>{\raggedright\let\newline\\\arraybackslash\hspace{0pt}}m{#1}}
\newcolumntype{C}[1]{>{\centering\let\newline\\\arraybackslash\hspace{0pt}}m{#1}}
\newcolumntype{R}[1]{>{\raggedleft\let\newline\\\arraybackslash\hspace{0pt}}m{#1}}

\makeatother

\begin{document}

\title{A quantum-chemical perspective on the \\ laser-induced alignment and orientation 
dynamics \\ of the CH$_3$X (X = F, Cl, Br, I) molecules 
}

\author{Irén Simkó}
\thanks{Equal contribution.}
\affiliation{Institute of Chemistry, ELTE Eötvös Loránd University and ELKH-ELTE
Complex Chemical Systems Research Group, H-1117 Budapest, Pázmány
Péter sétány 1/A, Hungary}

\author{Kalyani Chordiya}
\thanks{Equal contribution.}
\affiliation{ELI-ALPS, ELI-HU Non-Profit Ltd., Dugonics tér 13 and \\
University of Szeged,  D\'om t\'er 9, H-6720 Szeged, Hungary}

\author{Attila G. Császár}

\affiliation{Institute of Chemistry, ELTE Eötvös Loránd University and ELKH-ELTE
Complex Chemical Systems Research Group, H-1117 Budapest, Pázmány
Péter sétány 1/A, Hungary}

\author{Mousumi Upadhyay Kahaly}
\email{mousumi.upadhyaykahaly@eli-alps.hu}

\affiliation{ELI-ALPS, ELI-HU Non-Profit Ltd., Dugonics tér 13 and \\
University of Szeged,  D\'om t\'er 9, H-6720 Szeged, Hungary}

\author{Tamás Szidarovszky}
\email{tamas.janos.szidarovszky@ttk.elte.hu}

\affiliation{Institute of Chemistry, ELTE Eötvös Loránd University and ELKH-ELTE
Complex Chemical Systems Research Group, H-1117 Budapest, Pázmány
Péter sétány 1/A, Hungary}

\date{\today}
\begin{abstract}
Motivated by recent experiments, the laser-induced alignment-and-orientation (\AnO)
dynamics of the prolate symmetric top CH$_3$X (X = F, Cl, Br, I) molecules
is investigated, with particular
emphasis on the effect of halogen substitution on the
rotational constants, dipole moments, and polarizabilities of these species, 
as these quantities determine the \AnO\ dynamics. 
Insight into possible control schemes for preferred \AnO\ dynamics 
of halogenated molecules and
best practices for \AnO\  simulations are provided, as well.
It is shown that for accurate \AnO-dynamics simulations it is necessary to 
employ large basis sets and high levels of electron correlation when computing
the rotational constants, dipole moments, and polarizabilities.
The benchmark-quality values of these molecular parameters, corresponding to the 
equilibrium, 
as well as the vibrationally averaged 
structures are 
obtained with the help of the
focal-point analysis (FPA) technique and explicit electronic-structure computations
utilizing the gold-standard CCSD(T) approach, basis sets up to
quintuple-zeta quality, core-correlation contributions and,
in particular, relativistic effects for CH$_3$Br and CH$_3$I. 
It is shown that the different \AnO\ behavior of the CH$_3$X molecules in the optical regime is 
mostly caused by the differences in their polarizability anisotropy,
in other terms, the size of the halogen atom.
In contrast, the \AnO\ dynamics of the CH$_3$X series induced by an intense 
few-cycle THz pulse is mostly governed 
by changes in the rotational constants, due to the similar dipole moments
of the CH$_3$X molecules.
The \AnO\ dynamics is most sensitive to the 
$B$ rotational constant:
even the difference between its equilibrium and vibrationally-averaged values
results in noticeably different \AnO\  dynamics.
The contribution of rotational states having different symmetry, 
weighted by nuclear-spin statistics, 
to the \AnO\ dynamics is also studied.
\end{abstract}
\maketitle

\begin{spacing}{1.125}
\tableofcontents
\end{spacing}

\section{Introduction}\label{Introduction}
Laser-induced molecular rotational alignment and orientation (\AnO) has been investigated,
both theoretically and experimentally, for a number of decades. 
Different \AnO\ realizations have been proposed and demonstrated \cite{03StSe,10OhHa,13LeKrDoKa,19KoLeSu}, 
such as adiabatic \cite{08MaBr,16KoMi,99LaSaSaWe,95FrHe} and non-adiabatic \AnO\ \cite{04BiPoPeVi,16DaKaFl,18SoIwYaHa,01RoVr,99Se}, \AnO\ along one \cite{17ShChSoCh,19PiShChSt,18SoIwYaHa}, two \cite{16KoMi}, or three dimensions \cite{06LeViCoSt,12ArSe,14ReMaKu,00LaHaBjSt},
using linearly \cite{16DaKaFl,17ShChSoCh,18SoIwYaHa} or circularly \cite{19PiShChSt} polarized pulses,
employing single \cite{17ShChSoCh,19PiShChSt} or multiple pulses \cite{04BiPoPeVi,06LeViCoSt,14ReMaKu,16DaKaFl}, utilizing optical \cite{06LeViCoSt,16DaKaFl,17ShChSoCh} and/or terahertz (THz) \cite{16DaKaFl,21TuXuFiNe,11ZhFiNe} pulses,
optical centrifuges \cite{16KoMi}, alignment in helium droplets \cite{17ShChSoCh,19PiShChSt}, to name just a few developments. 
In addition to the various practical applications of \AnO\ in molecular sciences,
including, but not limited to 
rotational coherence spectroscopy (which can be used to determine accurate rotational constants \cite{92Felker,18ScLeSc,02Riehn,20ChChScJo}),
chemical reaction control, isotope and photofragment separation,
and molecule trapping, the appearance of attochemistry \cite{17NiDeCaPa,09KrIv,17KuDuKaMo}
has stimulated renewed interest in \AnO\ dynamics.
Since molecular rotations typically proceed on a picosecond to nanosecond timescale, 
efficient rotational \AnO\ allows for carrying out femtosecond or 
attosecond dynamics experiments in a molecule-fixed frame, providing 
potentially much richer information than experiments carried out on isotropic samples
\cite{15KrMiBaRu,09BiClWuLe,09MaMaViJo}.
Furthermore, coherent molecular rotations can be exploited to provide unique media 
for high-harmonics generation \cite{16SkChCaSa,16FaPrHeBi,21YaLiZhTu}.

The photochemistry of various methyl halide molecules, that is 
CH$_3$F, CH$_3$Cl, CH$_3$Br, and CH$_3$I, has drawn attention
due partially to the molecules' impact on atmospheric science. 
These halogen-containing molecules have a simple structure 
(they are symmetric tops) and their interaction with laser fields
has often been studied \textit{via} different experimental and 
theoretical methods.
For example, Hamilton \textit{et al}. \cite{05HaSeEjPo} studied, both experimentally 
and theoretically, nonadiabatic alignment of CH$_3$I,
focusing on the shape and intensity of the alignment revivals, 
including their dependence on the pulse duration and their behavior at long times,
where centrifugal distortion effects become important.
Recent experiments have been performed by 
Luo \textit{et al}. \cite{17LuHuYuZh} to investigate the \AnO\ dynamics 
of thermally averaged and quantum-state-selected CH$_3$I molecules, 
employing intense femtosecond 800 nm laser fields.
The same group investigated how the laser-induced alignment affects the double ionization and chemical bond rearrangement in CH$_3$Cl \cite{18LuZhHuYu}.
In another study,
He \textit{et al}. \cite{16HePaYaLu} measured the effect of 
laser-induced alignment on the ion yields of quantum-state-selected
CH$_3$I and CH$_3$Br molecules, in the 10--1000 TW\,cm$^{-2}$ 
intensity regime for the ionization laser. 
The angle-dependent strong-field ionization of halomethanes was studied by
S\'andor \textit{et al}. \cite{19SaSiMaGo}, who employed short and intense 
optical laser pulses to align the molecules and revealed that even though 
their electronic structure has similar character,
CH$_3$F, CH$_3$Cl, and CH$_3$Br ionize \textit{via} different mechanisms.

Recently, in order to aid experimentalists carrying out such simulations, 
a user-friendly and cross-platform software was developed, called LIMAO \cite{18SzJoYa},
where the acronym stands for ``laser-induced molecular alignment and orientation''. 
Although running \AnO\ dynamics simulations with LIMAO is not difficult, 
the program does require, as all \AnO\ simulations, molecular parameters that 
might not be readily available in the literature.
In such cases the most straightforward approach is to compute these parameters
using standard quantum-chemical (more precisely, electronic-structure) 
techniques \cite{00HeJoOl}.
The main motivation of this work was to 
identify the quantum-chemical techniques suitable for obtaining accurate \AnO-related
molecular parameters, and to 
study how halogen substitution affects \AnO\ dynamics through these parameters. We
also investigate, in a practical sense, some general features
of running \AnO\ simulations. 
\raggedbottom

\section{\label{Theory}Theory}
During this study, the laser-induced rotational dynamics was simulated by directly solving 
the appropriate time-dependent Schr\"odinger equation (TDSE), as implemented in
the code LIMAO \cite{18SzJoYa}.
In short, LIMAO utilizes a field-free eigenstate basis to solve the TDSE  
    \begin{equation}
        {\rm i}\hbar \partial_t \vert \Psi (t) \rangle = \hat{H}(t) \vert \Psi (t) \rangle,
        \label{eq:TDSE}
    \end{equation}
based on the time-dependent Hamiltonian $\hat{H}(t)$ of a molecule interacting with 
an external electric field. 
By treating electronic excitation perturbatively, $\hat{H}(t)$ can be written \cite{15HaOh} as
\begin{equation}
    \hat{H}(t)=
    \hat{H}_0-\bm{\varepsilon}(t)\hat{\bm{\mu}}-\frac{1}{2}\bm{\varepsilon}(t)(\hat{\bm{\alpha}}\bm{\varepsilon}(t)),
\label{eq:hamiltonian}
\end{equation}
where $\hat{H}_0$ is the field-free rotational Hamiltonian, $\hat{\bm{\mu}}$ and 
$\hat{\bm{\alpha}}$ are the permanent molecular dipole and polarizability operators,
respectively, and $\bm{\varepsilon}(t)$ is the three-dimensional (3D) external 
electric-field vector.
Assuming a linearly polarized external field, {\it i.e.}, 
$\bm{\varepsilon}(t)=(0,0,\varepsilon(t))$, and using the $\vert \Psi ^{JMn} \rangle$ 
field-free eigenstates, satisfying the time-independent Schr\"odinger equation
    \begin{equation}
        \hat{H}_0 \vert \Psi ^{JMn} \rangle = E ^{Jn} \vert \Psi ^{JMn} \rangle,
    \label{eq:field_free_schr_Eq}
    \end{equation}
one can construct the matrix representation of Eq.~(\ref{eq:hamiltonian}), 
yielding the matrix elements
    \begin{equation}
    \begin{split}
        \langle \Psi ^{JMn} \vert \hat{H}(t) \vert \Psi ^{J'M'n'} \rangle = \\ E ^{Jn} \delta_{JJ'} \delta_{nn'} \delta_{MM'}-\langle \Psi ^{JMn} \vert \bm{\varepsilon}(t)\bm{\mu} \vert \Psi ^{J'M'n'} \rangle -\frac{1}{2} \langle \Psi ^{JMn} \vert \bm{\varepsilon}(t)(\bm{\alpha}\bm{\varepsilon}(t)) \vert \Psi ^{J'M'n'} \rangle = \\ E ^{Jn} \delta_{JJ'} \delta_{nn'} \delta_{MM'}- \\ \varepsilon(t) \sum_{k=-1}^{1} \langle \Psi ^{JMn} \vert {D_{0k}^{1}}^*  \mu ^{{\rm BF},(1,k)} \vert \Psi ^{J'M'n'} \rangle - \\ \frac{\varepsilon^2(t)}{\sqrt{6}} \Bigg[ \sum_{k=-2}^{2} \langle \Psi ^{JMn} \vert {D_{0k}^{2}}^*  \alpha ^{{\rm BF},(2,k)} \vert \Psi ^{J'M'n'} \rangle - \frac{1}{\sqrt{2}}\langle \Psi ^{JMn} \vert \alpha ^{{\rm BF},(0)} \vert \Psi ^{J'M'n'} \rangle \Bigg].
    \end{split}
    \label{eq:general_hamiltonian_in_eigenstate_basis}
    \end{equation}
In Eqs.~(\ref{eq:field_free_schr_Eq}) and (\ref{eq:general_hamiltonian_in_eigenstate_basis}),
$J$ is the rotational angular-momentum quantum number, quantum number $M$ refers to
the projection of the angular momentum onto the space-fixed $z$ axis, 
$n$ represents all the other quantum numbers,
and $E ^{Jn}$ are field-free molecular eigenenergies.
In the same equations, $\mu ^{{\rm BF},(1,k)}$ and the
$\alpha ^{{\rm BF},(2,k)}$ and $\alpha ^{{\rm BF},(0)}$ pair are
the body-fixed molecular dipole and polarizability in the spherical-basis 
representation \cite{88Zare,06BuJe}, 
respectively, for which transformation to the space-fixed components can be carried out 
\textit{via} the Wigner-D matrices $D_{0k}^{i}$ \cite{88Zare,06BuJe}. 
The specific form of $\vert \Psi ^{JMn} \rangle$ and the matrix elements of 
Eq.~(\ref{eq:general_hamiltonian_in_eigenstate_basis}) depend on the type of molecule 
(rotational top) considered; for details, see Ref. \citenum{18SzJoYa}. 
In this work only symmetric tops are investigated; 
therefore, $\vert \Psi ^{JMn} \rangle$ are the symmetric top eigenfunctions $\vert J K M \rangle$,
where $K$ is the quantum number corresponding to the projection of the 
angular momentum onto the body-fixed $z$ axis.
For a given initial wave function, $\vert \Psi ^{JMn} \rangle$, 
and a specific external electric field, $\bm{\varepsilon}(t)$,
the TDSE is solved by numerical propagation, employing the Hamiltonian matrix elements given in Eq.~(\ref{eq:general_hamiltonian_in_eigenstate_basis}).

LIMAO allows simulations at finite rotational temperatures under the assumption that 
the initial population in the different rotational eigenstates 
satisfies the Boltzmann distribution.
The population of the \textit{i}th eigenstate, $ P_{i}$, at thermal equilibrium is then given by 
\begin{equation}
\label{eq: Population of rot. st.}
    P_{i} = \frac{g_ie^{- \frac{E_{i}}{k_{\rm B}T}}}{Q_{\rm rot}(T)},
\end{equation}
where $Q_{\rm rot}(T)=\sum_{l} g_le^{- \frac{E_{l}}{kT}}$ is the rotational partition function,
$k_{\rm B}$ is Boltzmann's constant, $E_{i}$ is the eigenenergy of the $i$th rotational state,
and $g_{i}$ stands for the nuclear spin statistical weight (NSSW) of the 
\textit{i}th rotational eigenstate \cite{06BuJe}.
The occurrence of NSSWs in Eq.~(\ref{eq: Population of rot. st.}) 
is a direct consequence of the Pauli exclusion principle \cite{06BuJe}. 
In brief, rotational eigenstates can only be paired with nuclear spin functions
for which the product rotational-nuclear spin wave function has proper 
nuclear permutation symmetry, \textit{i.e.}, it changes sign (remains unchanged)
to the permutation of identical fermionic (bosonic) nuclei. 
NSSW 
plays an important role in the relative populations of different rotational states 
at finite temperatures. 
The NSSWs of rotational states can be determined following the procedure described 
in the supplementary material of Ref.~\citenum{18SzJoYa}.

The temporal evolution of the expectation value of a physical quantity $\hat{A}$, 
$\langle\hat{A}\rangle(t)$, is then expressed as
\begin{equation}
\label{eq: temp. evol. of expectation value}
    \langle\hat{A}\rangle(t) = \frac{1}{Q_{\rm rot}(T)} \sum_{i} \bra{\Psi^{(i)}(t) }\hat{A} \ket{\Psi^{(i)}(t)} g_{i}e^{- \frac{E_{i}}{kT}},
\end{equation}
where $\Psi^{(i)}(t)$ is the time-dependent rotational wave packet 
when the initial condition is set to be the \textit{i}th rotational eigenstate;
\begin{equation}
\label{eq:wavepacket}
    \Psi^{(i)}(t)=\sum_{JMn} C_{JMn}^{(J_i M_i n_i)}(t) \ket{\Psi^{JMn}},
\end{equation}
where $C_{JMn}^{(J_i M_i n_i)}(t)$ are the time-dependent expansion coefficients. Thus,
\begin{equation}
\bra{\Psi^{(i)}(t)} \hat{A} \ket{\Psi^{(i)}(t)} = \sum_{JMn, J'M'n'} C_{JMn}^{(J_i M_i n_i)^*}(t) C_{J'M'n'}^{(J_i M_i n_i)}(t) \bra{\Psi^{JMn}}\hat{A} \ket{\Psi^{J'M'n'}},  
\end{equation}
with $C_{JMn}^{(J_i M_i n_i)}(t = 0) = \delta_{J J_i}\delta_{M M_i}\delta_{n n_i}$. 
For quantifying the alignment, $\hat{A}=\cos^2(\theta)$ can be used, 
while for the orientation it is appropriate to utilize $\hat{A}=\cos(\theta)$,
where $\theta$ is the angle between the laboratory-fixed and body-fixed $z$ axes.
The specific form of  the $\bra{\Psi^{JMn}}\cos(\theta) \ket{\Psi^{J'M'n'}}$ and
$\bra{\Psi^{JMn}}\cos^2(\theta) \ket{\Psi^{J'M'n'}}$ matrix elements depend on the
type of molecule considered; see Ref. \citenum{18SzJoYa} for details. 
The matrix elements for $\cos^2(\theta)$ and the $\ket{JKM}$ symmetric-top eigenfunctions are
\begin{equation}
\label{eq:symmtop_align}
\begin{split}
    \braket{JKM|\cos^2(\theta)|J'K'M'}=\\
    \frac{1}{3}\delta_{JJ'}\delta_{KK'}\delta_{MM'}+\delta_{MM'}\frac{2}{3}\sqrt{(2J+1)(2J'+1)}(-1)^{M+K'}
    \begin{pmatrix}
       J & 2 & J' \\
       M & 0 & -M
    \end{pmatrix}
    \begin{pmatrix}
       J & 2 & J' \\
       K & 0 & -K'
    \end{pmatrix}
\end{split}
\end{equation} 
and
\begin{equation}
\label{eq:symmtop_orient}
    \braket{JKM|\cos(\theta)|J'K'M'}=
    \delta_{MM'}\sqrt{(2J+1)(2J'+1)}(-1)^{M+K'}
    \begin{pmatrix}
       J & 1 & J' \\
       M & 0 & -M
    \end{pmatrix}
    \begin{pmatrix}
       J & 1 & J' \\
       K & 0 & -K'
    \end{pmatrix},
\end{equation}
where $\big(:::\big)$
denotes the Wigner 3-$j$ symbols \cite{88Zare}.
It follows form the properties of the Wigner 3-$j$ symbols that 
$\braket{JKM|\cos^2(\theta)|J'K'M'}$ can be nonzero only if $M=M'$, $K=K'$, and 
$|J-J'|\leq 2$. 
Similarly, $\braket{JKM|\cos(\theta)|J'K'M'}$ can be nonzero only if
$M=M'$, $K=K'$, and $|J-J'|\leq 1$.

Depending on the temporal profile of the exciting pulse(s), one might achieve 
adiabatic or non-adiabatic, also called field-free, \AnO\ \cite{03StSe}. 
Adiabatic \AnO\ occurs when the external field changes much slower than the 
characteristic rotational timescale of the molecule, resulting in \AnO\ only during 
the presence of the external field. 
On the other hand, non-adiabatic \AnO\ can occur when the exciting pulse is 
shorter than the characteristic rotational timescale of the molecule, 
leading to the formation of a rotational wave packet, which might show,
under field-free conditions, \AnO\ after the external field subsides. 
In what follows we briefly describe how specific patterns appear in the temporal
evolution of field-free \AnO\ dynamics of molecules. 

Let us start with a qualitative description \cite{92Felker}. 
Initially, in the absence of an external field, the molecules have isotropic distribution,
the corresponding values of alignment and orientation are exactly 1/3 and 0, respectively.
Then, assuming a polarized laser pulse, a rotational wave packet is formed 
and the molecules are aligned or oriented parallel to the polarization axis. 
This arrangement, however, quickly disappears due to the 
dispersion
of the rotational wave packet and the sample shows a pseudoisotropic distribution. 
After a certain time rephasing occurs, causing a transient increase in the \AnO,
called the ``revival''.
This is followed again by 
dispersion
and the cycle starts over.

Now, let us turn to a more formal description \cite{92Felker}.
The laser pulse excites the sample and creates a rotational wave packet,
see Eq.~(\ref{eq:wavepacket}).
If the external field is turned off at $t_{\rm end}$, 
the time-dependent coefficients have the form  
\begin{equation}
C_{JMn}^{(J_i M_i n_i)}(t+t_{\rm end})=\exp\left(-\frac{\rm{i}}{\hbar}E^{Jn}t\right)C_{JMn}^{(J_i M_i n_i)}(t_{\rm end}),    
\end{equation}
corresponding to the field-free time evolution after $t_{\rm end}$. 
Then, the expectation value of $\hat{A}$ after the pulse is 
\begin{equation}
\label{eq:field-free_evol}
\langle\hat{A}\rangle(t+t_{\rm end})= \sum_{JMn, J'M'n'} C_{JMn}^{(J_i M_i n_i)^*}(t_{\rm end}) C_{J'M'n'}^{(J_i M_i n_i)}(t_{\rm end}) \bra{\Psi^{JMn}}\hat{A} \ket{\Psi^{J'M'n'}}\exp\left(-\frac{\rm i}{\hbar}\omega_{J'n',Jn}t\right),  
\end{equation} 
where $\omega_{J'n',Jn}=E^{J'n'}-E^{Jn}$. 
The time dependence of $\langle\hat{A}\rangle(t+t_{\rm end})$ is a result of 
the superposition of the exponential terms with different $\omega_{J'n',Jn}$ frequencies.
One could expect that the superposition never results in constructive interference, 
but this is not the case.
In the case of the rotational wave packet of linear or symmetric tops, 
the $\omega_{J'n',Jn}$ frequencies have the form 
\begin{equation}
    \omega_{J'n',Jn}=k_{J'n',Jn}\Omega
    \qquad\text{or}\qquad
    \omega_{J'n',Jn}=k_{J'n',Jn}\Omega+\Phi,
\end{equation}
where $k_{J'n',Jn}$ is an integer, while $\Omega$ and $\Phi$ are constants.
Therefore,  a constructive interference occurs in $\langle\hat{A}\rangle(t+t_{\rm end})$
if $t$ is an integer multiple of $h/\Omega$, where $h$ is Planck's constant,
leading to an observable ``revival'' \cite{92Felker}.

Next, we describe how revivals of symmetric-top molecules are manifested 
in the orientation and alignment. 
The matrix element of the orientation can be nonzero only if 
$K=K'$ and $|J-J'|\leq 1$ (see Eq. (\ref{eq:symmtop_orient})), 
so only these terms contribute to the superposition.
Therefore, using the $E^{J,K}=hcBJ(J+1)+hc(A-B)K^2$ energy formula of prolate symmetric tops,
\begin{equation}
    \omega_{J'K',JK}=E^{J'K'}-E^{JK}=
    \begin{cases}
2hcBJ+2hcB,\quad\text{if}\quad J'=J+1\text{ and } K'=K\\
0,\quad\text{if}\quad J'=J\text{ and }  K'=K,
\end{cases}
\end{equation}
so the $J'=J$ contributions are constant in time and the time periodicity of the 
orientation revivals is $T_{\rm rev}=1/(2Bc)$ due to the $J'=J+1$ contributions. 
In the case of alignment, the matrix element can be nonzero only if $K=K'$ and 
$|J-J'|\leq 2$ (see Eq. (\ref{eq:symmtop_align})).
Thus, 
\begin{equation}
    \omega_{J'K',JK}=E^{J'K'}-E^{JK}=
    \begin{cases}
4hcBJ+6hcB,\quad\text{if}\quad J'=J+2\text{ and } K'=K\\
2hcBJ+2hcB,\quad\text{if}\quad J'=J+1\text{ and } K'=K\\
0,\quad\text{if}\quad J'=J\text{ and }  K'=K.
\end{cases}
\end{equation}
In practice, the contributions from the different $J'=J+1$ cases cancel out.
Then, only the $J'=J+2$ terms contribute to the time evolution of the alignment, 
resulting in revivals with alternating polarity (half revival) and with $T_{\rm rev}=1/(4Bc)$
periodicity in time. 
Therefore, alignment revivals are expected to occur twice as often as orientation revivals.

\section{\label{ComputationalDetails}Computational details}
Details concerning the electronic-structure computations performed 
are described in this section. 
Readers who would like to skip this part should continue reading 
either at Subsection \ref{Substituent_effect}, where the relationship of halogen substitution
and the molecular parameters are explored, 
or at Section \ref{Single-optical-pulse alignment}, 
where the results of the laser-induced dynamics simulations are described.

The molecular parameters needed to solve Eq.~(\ref{eq:field_free_schr_Eq}) and to construct 
the matrix elements of Eq.~(\ref{eq:general_hamiltonian_in_eigenstate_basis}),
\textit{i.e.}, the rotational constants (calculated with average atomic masses), the dipole moments, and the polarizabilities,
were computed using either the CFOUR \cite{cfour} 
or the MOLPRO \cite{molpro} quantum-chemistry packages. 
The standard wave-function-theory quantum-chemical methods \cite{86HeRaScPo,00HeJoOl,06Jensen}
employed include the restricted Hartree--Fock (RHF) method \cite{51Roothaan}, 
its extension with electron correlation using second-order perturbation theory 
and the M{\o}ller--Plesset partitioning (MP2) \cite{34MoPe,80KrFrPo}, 
coupled-cluster (CC) theory \cite{66Cizek} 
with single and double excitations (CCSD) \cite{82PuBa}, and
the gold-standard CCSD(T) method \cite{89RaTrPoHe}, whereby CCSD is augmented with a 
perturbative correction for triple excitations.
We also performed computations within density-functional theory (DFT),
employing the popular B3LYP functional \cite{88Becke,93Becke,88LeYaPa}.

The atom-centered, fixed-exponent Gaussian basis sets utilized during the electronic-structure
computations have been developed by Dunning and co-workers \cite{89Dunning}.
The basis sets chosen include diffuse (aug) functions and occasionally
the core-valence correlation is also treated.
We use the following abbreviations throughout the manuscript: 
XZ = cc-pV$X$Z, aXZ = aug-cc-pV$X$Z, awcXZ = aug-cc-pwCV$X$Z, and 
awcXZpp = [aug-cc-pwC$X$Z on the H and C atoms and aug-cc-pwC$X$Z-pp on the halogen atom,
pp = pseudopotential], where $X$, the cardinal number of the bases, is either 
2(D), 3(T), 4(Q), or 5.
For the analysis of the computed quantum-chemistry data, 
we apply the focal-point analysis (FPA) approach \cite{93AlEaCs,98CsAlSc}. 
FPA utilizes the fact that the increments of electronic energies and properties 
on the basis set size and the level of electron correlation are more-or-less
independent from each other. 

All directly computed molecular parameters 
correspond to the equilibrium structure.
This becomes 
an issue when comparing directly computed (equilibrium) parameters 
to measured (effective) values, which are
usually expectation values in the ground vibrational state \cite{06Jensen}.
Thus, besides the equilibrium values of the rotational constants 
($A_{\rm e}$ and $B_{\rm e}$) and dipole moments ($\mu_{\rm e}$), 
their vibrationally averaged values ($A_0$, $B_0$, and $\mu_0$) were 
also computed for CH$_3$F and CH$_3$Cl at the
CCSD(T)\textunderscore FC/aug-cc-pVTZ level, using the relative atomic weights of 
$^1$H, $^{12}$C, $^{19}$F, and $^{35}$Cl. 
The accurate $A_0-A_{\rm e}$, $B_0-B_{\rm e}$, and $\mu_0-\mu_e$ differences can be
used as corrections that can be added to equilibrium parameter values computed at 
different levels of electronic-structure theory.
Vibrational corrections to the polarizability were not computed,
because the expected effect is much smaller than the uncertainty in the experimental
intensity values, 
most \AnO\ simulations do not require
very high accuracy for the polarizability
(in contrast, laser-induced \AnO\ dynamics is quite sensitive to the accuracy 
of the rotational constants (\textit{vide infra})). 
Highly accurate polarizability values that can be computed 
might become important in the future when the experimental intensity 
will be known more precisely.

During the \AnO\ dynamics simulations employing the LIMAO software, 
convergence of the simulated results with respect to the number of 
$\vert J K M \rangle$ rotational basis functions and the cutoff value
for the Boltzmann populations considered was ensured by gradually increasing 
and decreasing their values, respectively. 
An additional factor influencing \AnO\  dynamics, 
through the populations of different rotational states and the rotational
partition function, is the nuclear spin statistical weights (NSSW) 
of the different rotational levels. 
The CH$_3$X species are symmetric tops;
thus, they belong to the D$_\infty$ rotational symmetry group \cite{06BuJe}. 
The NSSWs for the irreps of the D$_\infty$ rotational 
symmetry group can be determined as prescribed by Bunker and Jensen \cite{06BuJe} 
or the supplementary material of Ref. \cite{18SzJoYa}. 
For all the CH$_3$X molecules considered, they are as follows: 
NSSW$^{\Sigma^+}$ = 2, NSSW$^{\Sigma^-}$ = 2, NSSW$^{E_{1}}$ = 1, 
NSSW$^{E_{2}}$ = 1, and NSSW$^{E_{3}}$ = 2. 
These NSSWs reflect the relative abundance of the rotational states 
belonging to different irreps, which has to be considered when computing the
thermal average of physical quantities, see Eq.~(\ref{eq: temp. evol. of expectation value}).
Note that the situation becomes different if different isotopologues are studied.
Even though they have very similar electronic properties, 
different isotopologues should be treated as different species 
in terms of \AnO\ dynamics, because the rotational constants and the NSSWs are different.

\section{On molecular parameters important for \AnO\ dynamics}
\subsection{Effects of electron correlation and basis set size on molecular parameters}
When the molecular parameters used in Eq. (\ref{eq:general_hamiltonian_in_eigenstate_basis}) are not available in the literature, their values can be computed with quantum-chemical methods.
By using different Gaussian basis sets and electron-correlation methods in the 
electronic-structure computations, and by utilizing the principles of the FPA
method \cite{93AlEaCs,98CsAlSc}, the accuracy of the different computed molecular parameters 
and their impact on the dynamics calculations can be estimated.
The CH$_{3}$X molecules chosen belong to the C$_{3v}$(M) molecular symmetry group \cite{06BuJe}
and their equilibrium structures possess C$_{3v}$ point-group symmetry.
Thus, they are symmetric-top molecules, 
for which second-order tensorial properties can be divided into a parallel 
and perpendicular component, both with respect to the molecular symmetry axis.
Tables \ref{table:FPA_for_structure} and \ref{table:FPA_for_mu_and_alpha} 
summarize the equilibrium molecular parameters most relevant for \AnO\ 
simulations, computed at various levels of sophistication.

\begin{table}[t!]
\renewcommand{\arraystretch}{1.25}
\caption{\small Computed 
equilibrium rotational constants, 
$A_{\rm e}$ and $B_{\rm e}$, 
and their method-dependent increments for the prolate symmetric-top CH$_3$X 
(X = F, Cl, Br, I) species, all given in cm$^{-1}$, obtained at different levels of
electronic-structure theory.}
\vspace{-0.8\baselineskip} 
\begin{center}
\begin{tabularx}{\textwidth}{c *{6}{Y}}
\hline \hline

\multicolumn{2}{c}{}	&\multicolumn{2}{c}{CH$_3$F}	&&	\multicolumn{2}{c}{CH$_3$Cl}	\\
method$^a$	&~~~basis$^a$	&~~$A_{\rm e}$	&~~$B_{\rm e}$	&&	$A_{\rm e}$	&~~$B_{\rm e}$	\\ \hline
RHF~	&	DZ	&	5.2602	&	0.8762	&&	5.2439	&	0.4410	\\
	& $\delta$aDZ	&	$-0.0155$	&	$-0.0077$&	&	0.0008	&	$-0.0010$	\\
	& $\delta$aTZ	&	0.0860	&	0.0152	&&	0.0896	&	0.0037	\\
	& $\delta$aQZ	&	0.0073	&	0.0015&	&	0.0090	&	0.0015	\\
	& $\delta$a5Z	&	0.0013	&	0.0003	&&	0.0030	&	0.0009	\\
$\delta$[MP2]	&	aQZ	&	$-0.0375$	&	$-0.0237$&	&	$-0.0416$	&	0.0041	\\
$\delta$[CCSD]	&	aQZ	&	$-0.0126$	&	0.0057	&&	$-0.0102$	&	$-0.0028$	\\
$\delta$[CCSD(T)]	&	aQZ	&	$-0.0208$	&	$-0.0066$&	&	$-0.0229$	&	$-0.0023$	\\
CCSD(T)	& $\delta$awcQZ	&	$-0.0070$	&	$-0.0014$	&&	0.0040	&	0.0007	\\
CCSD(T)	&	awcQZ	&	5.2602	&	0.8592	&&	5.2725	&	0.4448	\\ \hline
B3LYP	&	a5Z	&	5.2428	&	0.8516	&&	5.2644	&	0.4394	\\ \hline \hline
	\multicolumn{2}{c}{}		&	\multicolumn{2}{c}{CH$_3$Br}		&&	\multicolumn{2}{c}{CH$_3$I}			\\
method$^a$	&~~~basis$^a$	&~~$A_{\rm e}$	&~~$B_{\rm e}$	&&	$A_{\rm e}$	&~~$B_{\rm e}$	\\ \hline
CCSD(T)	&	awcQZ	&	5.2497	&	0.3210&	&		&		\\
	&	awcQZpp	&		&		&&	5.2371	&	0.2531	\\ \hline \hline

\end{tabularx}
\label{table:FPA_for_structure}
\end{center}
\vspace{-0.75\baselineskip} 
\begin{spacing}{1.125}
$^a$ `Method' gives the electronic-structure technique used for the computation
of the rotational constants.
Increments are represented by the symbol $\delta$. 
When in front of a method given in brackets, $\delta$ refers to the increment, 
obtained with the basis set specified under `basis', with respect to the method preceding 
the indicated method in the following sequence: RHF $\rightarrow$ MP2 $\rightarrow$ CCSD 
$\rightarrow$ CCSD(T). 
When $\delta$ is in front of a basis set, the numerical values show 
the increment with respect to the basis set in the preceding line of the table.
For the meaning of the basis-set abbreviations, see the text.
\end{spacing}
\end{table}

As Table \ref{table:FPA_for_structure} demonstrates, accurate determination 
of the molecular structures requires geometry optimizations carried out utilizing
extensive basis sets as well as high-level treatment of electron correlation. 
For example, the basis-set-size increments in the equilibrium rotational constants 
moving from aug-cc-pVDZ (aDZ) to aug-cc-pVTZ (aTZ) at the RHF level is 
comparable to the correlation effect introduced by MP2.
The next largest increments are from the more complete treatment of electron correlation
at the CCSD and CCSD(T) levels, and increasing the basis set aTZ to aug-cc-pVQZ (aQZ).
Accounting for core correlation, with the aug-cc-pwCVQZ (awcQZ) basis,
seems to be necessary only 
to achieve very high accuracy.
DFT (B3LYP) results are also included in Table \ref{table:FPA_for_structure}, 
showing that DFT, for this set of molecules and with the B3LYP functional,
with a large basis set is a viable, inexpensive alternative to the sophisticated 
\textit{ab initio} (wave-function-theory) methods if moderate accuracy suffices. 
In the case of CH$_3$F, the $A_0-A_{\rm e}$ and $B_0-B_{\rm e}$ vibrational corrections are $-0.07688$ 
and $-0.00813$ cm$^{-1}$, respectively. 
For CH$_3$Cl, the $A_0-A_{\rm e}$ and $B_0-B_{\rm e}$ corrections are  
$-0.06692$ cm$^{-1}$ and $-0.00382$ cm$^{-1}$, respectively.
These values are larger than the $\delta$[CCSD(T)] corrections.

\begin{table}[b!]
\renewcommand{\arraystretch}{1.25}
\caption{\small {Computed equilibrium molecular dipole ($\mu_e$) 
and static polarizability ($\alpha_{\parallel}$ and $\alpha_{\perp}$) values 
and their increments for the CH$_3$X species, all in atomic units.}}
\vspace{-0.85\baselineskip} 
\begin{center}
\begin{tabularx}{\textwidth}{c *{8}{Y}}
\hline \hline

	\multicolumn{2}{c}{}		&	\multicolumn{3}{c}{CH$_3$F}				&		\multicolumn{3}{c}{CH$_3$Cl}					\\
method$^a$	&	basis$^a$	& $\mu_e$	&	$\alpha_{\parallel}$	&	$\alpha_{\perp}$ 	&	$\mu_e$ 	&	$\alpha_{\parallel}$ 	&	$\alpha_{\perp}$ 	\\ \hline
RHF	&	awcDZ	&	0.8107	&	16.4172	&	15.1131	&	0.8421	&	34.3168	&	24.1080	\\
& $\delta$awcTZ	&	$-0.0051$	&	0.1448	&	0.2742	&	$-0.0126$	&	0.5703	&	0.8419	\\
	& $\delta$awcQZ	&	$-0.0017$	&	0.0113	&	0.0602	&	$-0.0020$	&	0.0970	&	0.2324	\\
	& $\delta$awc5Z	&	$-0.0001$	&	$-0.0082$	&	0.0053	&	0.0000	&	0.0111	&	$-0.0048$	\\
$\delta$[MP2]	&awcQZ	&	$-0.0628$	&	1.2168	&	0.8114	&	$-0.0659$	&	0.7762	&	0.8749	\\
$\delta$[CCSD(T)]	&	awcQZ	&	$-0.0044$	&	$-0.1128$	&	$-0.1028$	&	$-0.0070$	&	$-0.2378$	&	$-0.1372$	\\
CCSD(T)	&	awcQZ	&	0.7367	&	17.6773	&	16.1560	&	0.7546	&	35.5225	&	25.9200	\\ \hline
B3LYP	&	a5Z	&	0.7330	&	18.360	&	16.853	&	0.7701	&	37.036	&	26.899	\\ \hline \hline
	\multicolumn{2}{c}{}		&	\multicolumn{3}{c}{CH$_3$Br}					&	\multicolumn{3}{c}{CH$_3$I}					\\ 
method$^a$	&	basis$^a$	&	$\mu_e$ 	&	$\alpha_{\parallel}$ 	&	$\alpha_{\perp}$ 	&	$\mu_e$ 	&	$\alpha_{\parallel}$ 	&	$\alpha_{\perp}$ 	\\ \hline
CCSD(T)	&	awcTZ	&	0.7266	&	44.4575	&	31.8600	&		&		&		\\
	& $\delta$awcQZ	&	0.0144	&	$-0.0450$	&	$-0.0150$	&		&		&		\\
CCSD(T)	&	awcQZpp	&		&		&		&	0.6592	&	58.9875	&	43.1250	\\
$\delta$[CCSD(T)+MVD1]	&	awcTZ	&	$-0.0174$	&	0.0400	&	0.0100	&		&		&		\\
	&	awcQZ	&	$-0.0193$	&	0.0775	&	0.0600	&		&		&		\\
	&	awcQZpp	&		&		&		&	0.0003	&	0.0175	&	0.0075	\\
CCSD(T)+MVD1	&	awcQZ	&	0.7218	&	44.4900	&	31.9050	&		&		&		\\
	&	awcQZpp	&		&		&		&	0.6595	&	59.0050	&	43.1325	\\ \hline \hline

\end{tabularx}
\label{table:FPA_for_mu_and_alpha}
\end{center}
\vspace{-0.85\baselineskip} 
$^a$ `Method' gives the electronic-structure technique used during the computation.
Increments are represented by the $\delta$ symbol. 
When in front of a method given in brackets, $\delta$ refers to the increment, 
obtained with the basis set specified under `basis', with respect to the method preceding 
the indicated method in the following sequence:
RHF $\rightarrow$ MP2 $\rightarrow$ CCSD(T) and CCSD(T) $\rightarrow$ CCSD(T)+MVD1.
When the $\delta$ symbol is in front of a basis set, the numerical values show 
the increment with respect to the basis set in the preceding line of the table.
For the meaning of the basis-set abbreviations, see the text.
\end{table}

Table \ref{table:FPA_for_mu_and_alpha} shows that, similar to that of the molecular structure,
the most accurate dipole and polarizability values are obtained if 
large basis sets and theoretical methods accounting for a substantial part of
electron correlation are employed. This is due to the fact that the anisotropic 
charge distribution around the halogen nuclei play a crucial role in such molecules.
The dipole-moment values seem to be much more sensitive to electron correlation
than to basis-set size: the MP2 increment is more than ten(five) times larger
than the increment from aug-cc-pwCVDZ (awcDZ) to aug-cc-pwCVTZ (awcTZ) 
at the RHF level for CH$_3$F(CH$_3$Cl). 
Nevertheless, going beyond the MP2/awcTZ level seems to be necessary only if 
accuracy beyond two digits is required. 
The polarizability values are, however, much more sensitive to the size of the Gaussian basis.
In these cases, the basis set increment from awcDZ to awcTZ is comparable to the MP2 increment,
while the increment from awcTZ to awcQZ is comparable to the CCSD(T) increment. 
B3LYP with a large basis gives an accurate dipole for CH$_3$F;
however, the dipole of CH$_3$Cl is significantly overestimated.
B3LYP also fails to deliver accurate polarizabilities, as expected \cite{06Jensen}.

\begin{table}[t!]
\renewcommand{\arraystretch}{1.35}
\caption{\small {Recommended 
molecular parameters important for \AnO\ processes: 
rotational constants 
($A_{\rm e}$, $B_{\rm e}$, $A_0$, and $B_0$),
dipole moments ($\mu_{\rm e}$ and $\mu_0$), and
polarizabilities ($\alpha_{\parallel}$, $\alpha_{\perp}$, and anizotropy,
$\Delta \alpha$) along with values \textit{of experimental origin}, when available,
given in parentheses below the results obtained within this study (CCSD(T)(+MVD1)/aug-cc-pwCVQZ(-pp) level, see text for details).
Additional atomic and molecular properties shown are the carbon-halogen bond length,
$r_{\rm C-X}$ (computed in this work), 
the van der Waals radius of the halogen atom, $r_{\rm X}$ \cite{mantina2009consistent},
the electronegativity, $EN_{\rm X}$ \cite{lide2004CRCHandbook}, 
the first ionization energy, $IE_{\rm X}$ \cite{lide2004CRCHandbook}, 
the electron affinity, $EA_{\rm X}$ 
\cite{lide2004CRCHandbook}, 
and the atomic polarizability of the halogen atom, $\alpha_{\rm X}$ \cite{schwerdtfeger2019table}.}
}
\vspace{-0.85\baselineskip} 
\begin{center}
\begin{tabular}{lcccc}
\hline \hline
 
                        	&	 CH$_3$F 	&	 CH$_3$Cl 	&	 CH$_3$Br 	&	CH$_3$I	\\	\hline
$A_{\rm e}$/cm$^{-1}$ &	5.260\,2	&	5.272\,5	&	5.249\,7	&	5.237\,1	\\	
                            &	(5.265\,00)$^a$ &	(5.267\,80)$^b$ &	(5.246\,31)$^c$ &	(5.228)$^d$	\\
$B_{\rm e}$/cm$^{-1}$ &	0.859\,2	&	0.444\,8	&	0.321\,0	&	0.253\,1	\\	
                            &(0.859\,84)$^a$ &(0.447\,382)$^b$ &(0.321\,92)$^c$ &	(0.252\,35)$^d$  \\
$A_{\rm 0}$/cm$^{-1}$ &	5.183\,3	    &	5.205\,6     &	    &	    	\\	
                         	&	(5.182\,00)$^e$   &(5.205\,30)$^b$  &       &   		\\
$B_{\rm 0}$/cm$^{-1}$ &	0.851\,1	&	0.441\,0 	&	    	&	    	\\	
                         	& (0.851\,79)$^e$ & (0.443\,40)$^b$ &           &	        \\
$\mu_{\rm e}$/D            	&	1.872\,4	&	1.917\,9	&	1.834\,4	&	1.676\,2	\\
$\mu_{\rm 0}$/D    	&	1.858\,5	&	1.897\,2 	&	    	&	    	\\
                         	&	(1.85 $\pm$ 1\%)$^f$  &	(1.87 $\pm$ 1\%)$^f$  &	(1.81 $\pm$ 1\%)$^f$ &	(1.62 $\pm$ 2\%)$^f$  \\	
$\alpha_{\perp}$/\AA$^3$ 	&	2.394	&	3.841	&	4.728	&	6.392	\\	
$\alpha_{\parallel}$/\AA$^3$&	2.620	&	5.264	&	6.593	&	8.744	\\	
$\Delta \alpha$/\AA$^3$  	&	0.225	&	1.423	&	1.865	&	2.352	\\	\hline
$r_{\rm C-X}$/\AA        	&	1.383	&	1.778	&	1.934	&	2.129	\\	
$r_{\rm X}$/\AA         	&	1.47	&	1.75	&	1.83	&	1.98	\\	
$EN_{\rm X}$             	&	3.98	&	3.16	&	2.96	&	2.66	\\	
$IE_{\rm X}$/eV            	&	17.423	&	12.968	&	11.814	&	10.451	\\	
$EA_{\rm X}$/eV            	&	3.401	&	3.613	&	3.364	&	3.059	\\	
$\alpha_{\rm X}$/\AA$^3$ 	&	0.554(12)	&	2.163(15) &	3.11(15)	&	4.88(19)	\\	\hline \hline

\end{tabular}
\end{center}
\label{table:molpropsummary}
\vspace{-0.4\baselineskip} 
Sources of the experimental rotational constants: 
$^a$ Ref. \citenum{99DeBrThPa}, $^b$ Ref. \citenum{81JeBrGu}, 
$^c$ Ref. \citenum{94Graner}, $^d$ Ref. \citenum{72MaOv},

$^e$ Ref. \citenum{93PaHsChPr}.

$^f$ Source of the dipole moment corresponding to the ground 
vibrational state: Ref. \citenum{67NeLiMa}. 
The experimental uncertainties reported include the variation due to the
different vibrational states and different measurement techniques; 
therefore, they should cover the equilibrium values, as well.
\end{table}

Relativistic effects might also be significant for heavy elements, because their 
inner electrons can have velocities comparable to the speed of light. 
Therefore, relativistic effects for the Br- and I-containing compounds were estimated
by computing one-electron mass-velocity–Darwin (MVD1) \cite{76CoGr,97Balasubr,01TaCsKlQu}
corrections.
All-electron calculations were possible for CH$_3$Br, 
showing MVD1 corrections comparable to the basis set increment from awcTZ to awcQZ,
affecting the dipole and polarizability values in their second 
and third digits, respectively. 
For CH$_3$I, the basis set describes core electrons with a pseudo potential,
implicitly containing relativistic effects. 
Therefore, only frozen-core calculations were carried out, and, naturally, 
very small MVD1 corrections were obtained.     

The $\mu_0-\mu_{\rm e}$ difference computed at the
CCSD(T)\textunderscore FC/aug-cc-pVTZ level is $-0.00548$ and $-0.00813$ a.u. 
for CH$_3$F and CH$_3$Cl, respectively. 
These values are similar to the $\delta$[CCSD(T)] correction.

Finally, following the FPA protocol, the complete-basis-set (CBS) limit
for the RHF dipole and polarizability values, as well as the CBS of 
their $\delta$[MP2] increment was extrapolated for all the CH$_3$X molecules. 
In all occurrences we found that the CBS correction with respect to the 
awcQZ(pp) basis set is orders of magnitude smaller than the $\delta$[CCSD(T)] increment;
therefore, the CBS values are not reported here, and the uncertainties of the best 
simulated values are estimated from above by the $\delta$[CCSD(T)] increment values.

\subsection{Substituent effects on chemical properties} \label{Substituent_effect}

Table~\ref{table:molpropsummary} summarizes the equilibrium molecular properties
computed as part of this work, obtained with the most sophisticated methods, 
CCSD(T) (+MVD1 for the dipole and polarizabilities of CH$_3$Br and CH$_3$I), and 
largest basis sets (awcQZ for CH$_3$F, CH$_3$Cl, and CH$_3$Br;  awcQZpp for CH$_3$I) 
employed for the given molecule and property. 
These values were employed in the LIMAO simulations. 
The vibrationally averaged rotational constants and dipole moments
of CH$_3$F and CH$_3$Cl were computed by adding the computed 
$A_0-A_{\rm e}$, $B_0-B_{\rm e}$, and $\mu_0-\mu_{\rm e}$ corrections to the 
equilibrium values. 
Table~\ref{table:molpropsummary} also contains
values of experimental origin, from the literature \cite{99DeBrThPa,81JeBrGu,94Graner,72MaOv,67NeLiMa,93PaHsChPr,66Herzberg},
along with some atomic properties taken also from the literature \cite{lide2004CRCHandbook,kandalam2015superhalogens}. 
As shown in Table \ref{table:molpropsummary},
there is a slight increase in the dipole moment going from CH$_{3}$F to CH$_{3}$Cl. 
On the other hand, due to the screening effect of the additional $d$ orbitals,
we see a decrease in the value of the dipole when moving from CH$_{3}$Cl to
CH$_{3}$I \cite{pritchard1955concept}. 
In the case of the bond lengths and the polarizability 
values, a steady increase is observed from CH$_{3}$F to CH$_{3}$I,
while the 
$B_{\rm e}$ rotational constant
decreases in this order and 
$A_{\rm e}$ shows only a minor variation, in agreement with the changes
in the structures of the prolate symmetric-top CH$_3$X species. 
As to the atomic properties, the electron affinity $EA_{\rm X}$ of the halogen atoms
follows the trend shown by the molecular dipole \cite{pritchard1955concept,kandalam2015superhalogens},
the van der Waals radius $r_{\rm X}$ increases from F to I, 
while the electronegativity, ionization energy, and polarizability
of the atom, $EN_{\rm X}$, $IE_{\rm X}$, and $\alpha_{\rm X}$, respectively, 
decrease in the same order.

\begin{table}[!b]
\renewcommand{\arraystretch}{0.80}
\caption{\small{Correlation of the polarizability anisotropy, $\Delta \alpha$, 
and the dipole moment, $\mu_{\rm e}$, with various atomic and molecular parameters, 
denoted by $\xi$, for the four CH$_3$X molecules (X = F, Cl, Br, and I).
$r_{\rm X}$ = van der Waals radius, $EN$ = electronegativity,
$IE$ = ionization energy, and $EA$ = electron affinity.
The correlation is denoted by  cor($\Delta \alpha$, $\xi$) and cor($\mu_{\rm e}$, $\xi$). 
The correlation of $X$ and $Y$ quantities is calculated as cor($X,Y$)=$\braket{X-\braket{X}}\braket{Y-\braket{Y}}/(\sigma_X\sigma_Y)$, 
where $\braket{}$ denotes the expectation value, and $\sigma_X$, 
$\sigma_Y$ denote the standard deviation of $X$ and $Y$, respectively.
The sample-size-adjusted correlation coefficient,
cor$^*$=sgn(cor)($\vert$cor$\vert-$0.95)/0.05 (where sgn is the sign function) 
is computed for the statistically significant correlations, 
meaning $\vert$cor$\vert>$0.95 (see text for details).
Large cor$^*$ values are boldfaced, indicating strong linear relationship. }}
\vspace{-0.85\baselineskip} 
    \label{tbl:correlation}
    \begin{center}
        \begin{tabular}{lrrrr}
        \hline \hline
        $\xi$ & cor($\Delta \alpha$, $\xi$) & ~~cor$^*$($\Delta \alpha$, $\xi$) & ~~cor($\mu_{\rm e}$, $\xi$) & ~~cor$^*$($\mu_{\rm e}$, $\xi$) \\ \hline
        $\mu_{\rm e}$  &$-0.666$ &  & 1.000 & 1.000 \\
        $\alpha_{\perp}$ & 0.965 & 0.302 & $-0.83$5 &  \\
        $\alpha_{\parallel}$ & 0.985 & 0.705 & $-0.782$ &  \\
        $\Delta \alpha$ & 1.000 & 1.000 & $-0.666$ &  \\
        $r_{\rm C-X}$ & 0.999 & \textbf{0.981} & -0.697 &  \\
        $r_{\rm X}$ & 1.000 & \textbf{0.992} & -0.682 &  \\
        $EN_{\rm X}$ & $-0.998$ & \textbf{--0.955} & 0.625 &  \\
        $IE_{\rm X}$ & $-0.997$ & \textbf{--0.931} & 0.604 &  \\
        $EA_{\rm X}$ & $-0.525$ &  & 0.977 & 0.538 \\
        $\alpha_{\rm X}$ & 0.968 & 0.368 & $-0.828$ &
        \\ \hline \hline
        \end{tabular}
    \end{center}
    \end{table}

Even though this work is limited to CH$_3$X molecules, 
our aim has been to arrive at conclusions as general as possible, in order to aid 
future work in designing and efficiently executing \AnO-related research.
We investigated, in this regard, the connection between molecular properties
relevant in \AnO\ processes and the chemical and physical properties 
of the test molecules or the atoms within. 
Thus, we calculated the correlation coefficient of $\Delta \alpha$ and the dipole moment
with the different atomic and molecular parameters (see Table \ref{tbl:correlation}).
The seemingly large correlation coefficients found can be misleading 
because the number of data points  is low (there are only four molecules),
so uncorrelated quantities can accidentally have large correlation coefficients. 
In the case of four data points, if the absolute value of the
correlation coefficient (cor) is larger than 0.95,
then the probability of an accidental correlation is only 5\% \cite{bevington2003data}.
Thus, we also calculated the sample-size-adjusted correlation coefficient, 
cor$^*$=sgn(cor)($\vert$cor$\vert-$0.95)/0.05 (where sgn is the sign function), 
to measure the correlation above the threshold value of 0.95, 
and indicated the large cor$^*$ values with boldfaced numbers in Table \ref{tbl:correlation}.

Based on the data of Table \ref{tbl:correlation},
one can see that there is a strong linear relationship between $\Delta \alpha$ 
and the carbon-halogen bond length ($r_{\rm C-X}$) and 
the van der Waals radius ($r_{\rm X}$), while there is a negative correlation 
with the electronegativity and the ionization energy. 
Polarizability is usually assumed to be nearly proportional to the molecular volume;
therefore, the linear dependency of $\Delta \alpha$,
the polarizability anizotropy, on the radius of the halogen atom
is somewhat surprising, especially in the light of the rigorous relation 
between the volume and the static polarizability of quantum systems revealed \cite{20SzGoChKa}. 
The linear dependency is probably caused by the very small sample size,
and one should not draw serious conclusions from this. Nonetheless, we find that the larger 
the molecular volume, the larger the polarizability.
For a set of molecules with similar,  non-spherical structure this translates to 
larger volumes leading to larger polarizability anisotropies.
The dipole moment shows significant correlation with the electron affinity
of the halogen. 
It is important to note that among the quantities investigated here, 
the dipole moment and the electron affinity are the only non-monotonous functions of 
the atomic number of the halogen.
Note that the dipole moment varies much less than the other chemical properties, 
probably because the partial charge, depending on the electronegativity difference between 
the carbon and the halogen atoms, decreases with the size of the halogen, 
but $r_{\rm C-X}$ increases, increasing the separation of the partial charges,
and these opposite effects tend to cancel each other in the dipole moment. 


\begin{figure*}[!b]
        \centering
        \includegraphics[width=\linewidth]{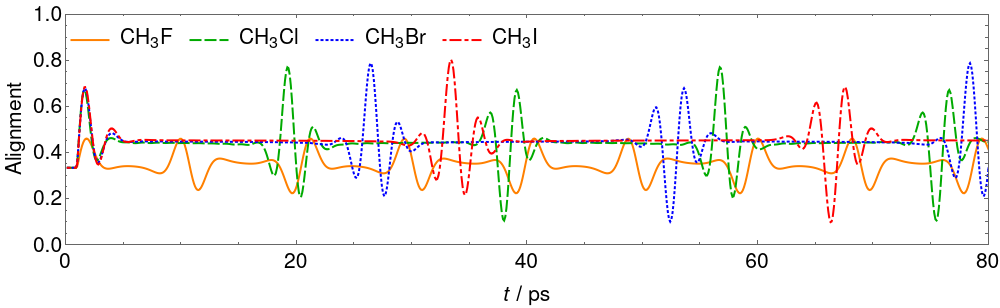}
        \\
        \includegraphics[width=\linewidth]{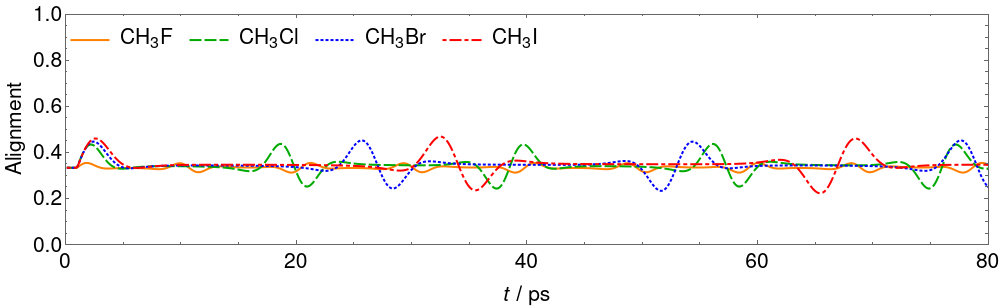}
        \caption{\small{Comparison of the laser-induced rotational alignment of the 
        CH$_3$X molecules for a medium- and a high-intensity optical pulse. 
        The pulse parameters chosen are as follows:
        $T$/K = 5, $\lambda$/nm = 800, $I/$TW\,cm$^{-2}$ = 30 (upper panel) 
        and 5 (lower panel), FWHM/fs = 100, and the pulse position is 1 ps.}}
        \label{fgr:CH3X_800nm_alignment}
        \end{figure*}

\section{Single-optical-pulse alignment} \label{Single-optical-pulse alignment}
In this section we are investigating the single-optical-pulse alignment dynamics
of the CH$_3$X species, for medium- to high-intensity (1$-$30 TWcm$^{-2}$) femtosecond pulses, 
which can be generated by tabletop Ti:sapphire setups \cite{00BrKr} or 
those available at dedicated institutes \cite{17KuDuKaMo}. 
We are studying the effect of molecular parameters, such as rotational constants, 
polarizability, and the contribution from different irreps of the rotation group
to the alignment, as well as the sensitivity of the alignment to the 
accuracy of these parameters. 
For the sake of completeness, 
we also investigate the effects of experimental conditions, \textit{i.e.},
the temperature and pulse parameters.

\subsection{The role of different molecular parameters}
First, let us investigate the effect of the change of substituent on \AnO\ dynamics
in the case of optical pulses ($\lambda=800$ nm is chosen). 
If the temperature and the laser parameters are kept fixed, 
the rotational constants and the polarizability anisotropy ($\Delta \alpha$)
determine the \AnO\ dynamics.
The rotational constants determine the time of the rotational revivals
and the maximal alignment: heavier molecules, with smaller 
$B_{\rm e}$ rotational constants, have 
longer rotational periods and smaller maximal alignment than the lighter molecules.
This is due to more initial rotational states being populated at the given temperature 
(\textit{vide infra}). 
Increasing the polarizability anisotropy increases the maximal alignment 
due to the higher degree of excitation, as shown below. 
For the CH$_3$X series, the rotational constants are decreasing with the size of the halogen,
while the polarizability anisotropy is increasing: 
$\Delta \alpha_{\rm CH_3F}<\Delta \alpha_{\rm CH_3Cl}<\Delta \alpha_{\rm CH_3Br}<\Delta \alpha_{\rm CH_3I}$,
with the largest difference found between X $=$ F and Cl. 
Thus, it is expected that increasing the size of the halogen increases 
the time between the revivals and increases the maximal alignment. 

For optical pulses, the dipole moment has no effect on \AnO\ dynamics,
because the field-dipole interaction averages out to zero during the fast 
oscillation of the field \cite{15HaOh}. This is also a reason why nonzero orientation
cannot be achieved using a single optical pulse. 
This was tested by comparing converged simulations explicitly including the 
dipole interaction with simulations utilizing the cycle-averaging 
approximation \cite{18SzJoYa, 15HaOh}, when the dipole interaction is neglected 
and only the envelope of the pulse is considered.
The two cases show identical alignment curves and no orientation.

\begin{figure*}[!b]
        \centering
        \includegraphics[width=\linewidth]{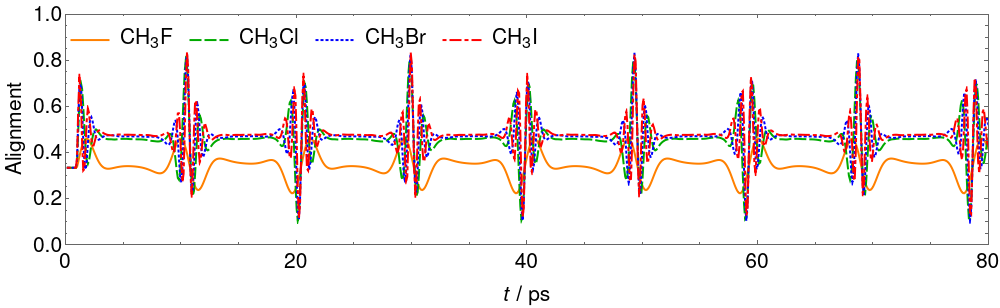}
        \\
        \includegraphics[width=\linewidth]{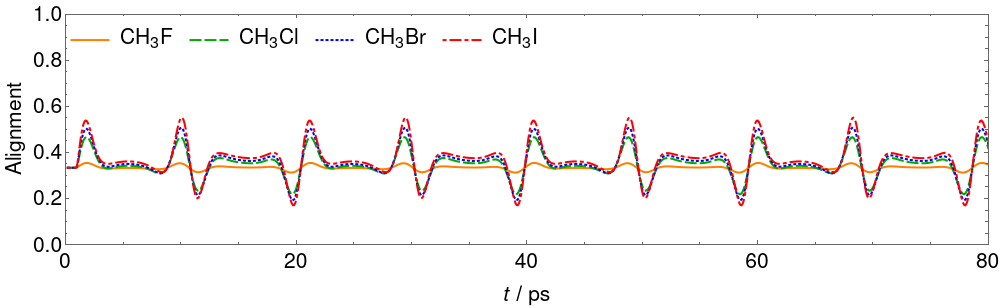}
        \caption{\small{Testing the effect of polarizability anisotropy on 
        laser-induced rotational alignment, by setting the rotational constant 
        to the corresponding value of CH$_3$F for all four molecules.
        The pulse parameters chosen are as follows:
        $T/$K = 5, $\lambda$/nm = 800, $I/$TW\,cm$^{-2}$ = 30 (upper panel) and
        5 (lower panel), FWHM/fs = 100, and the pulse position is 1 ps.}}
        \label{fgr:CH3X_800nm_rotconst}
\end{figure*}

Fig.~\ref{fgr:CH3X_800nm_alignment} shows the results computed for $T=5$ K
for two different pulse intensities, 5 TW\,cm$^{-2}$ (lower panel)
and 30 TW\,cm$^{-2}$ (upper panel), with a $\lambda=800$ nm Gaussian pulse 
positioned to 1 ps and having a full width half maximum (FWHM) of 100 fs. 
The alignment curves of the four molecules are clearly different, 
but similarities can also be identified.  
For the pulse with the larger intensity, the curve of CH$_3$F stands out 
with its lower base level and a reduced maximal alignment, 
while for the other three molecules the maximal alignments are quite similar, 
only the revival periods are different. 
We can achieve only a much smaller alignment with the lower intensity pulse,
and the revival peaks of CH$_3$F become smaller relative to the other three molecules.
The half-revival time periods, 
$T_{\rm rev}$ $=1/(4cB_{\rm e})$,
are 9.71, 18.75, 25.98, and 32.95 
for CH$_3$F, CH$_3$Cl, CH$_3$Br, and CH$_3$I, respectively. 
Based on the periodicity and the alternating polarity of the revivals, 
these are all $J$-type revivals \cite{92Felker}, 
where the molecule rotates around the 
$b$ and $c$ molecular axes.
When compared to experimental laser-induced alignment curves, 
our simulations exhibit agreement in terms of revival times and the 
shape of the revival curves; for examples, see Fig. 1 of 
Ref.~\citenum{18LuZhHuYu} for CH$_3$Cl, Fig. 3 of
Ref.~\citenum{16HePaYaLu} for CH$_3$Br, and 
Fig. 2 of Ref.~\citenum{05HaSeEjPo},
Fig. 4 of Ref.~\citenum{17LuHuYuZh} and 
Fig. 4 of Ref.~\citenum{16HePaYaLu} for CH$_3$I. 
Note that the exact measure of alignment and the experimental parameters are 
different from those in our simulations; therefore, 
the measured and computed alignment curves cannot be compared directly.

\begin{figure*}[!b]
            \centering
            \includegraphics[width=\linewidth]{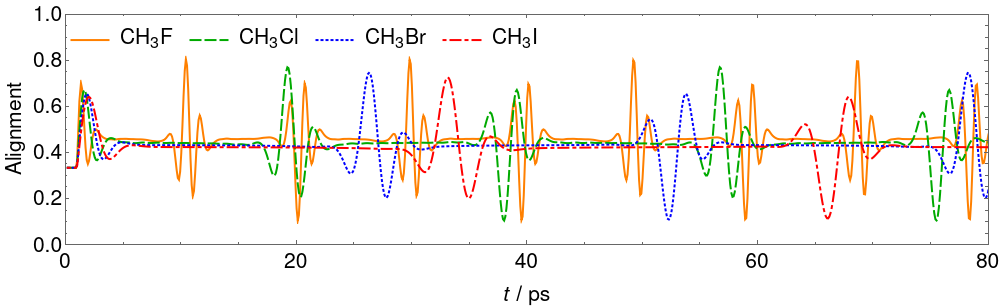}
            \\
            \includegraphics[width=\linewidth]{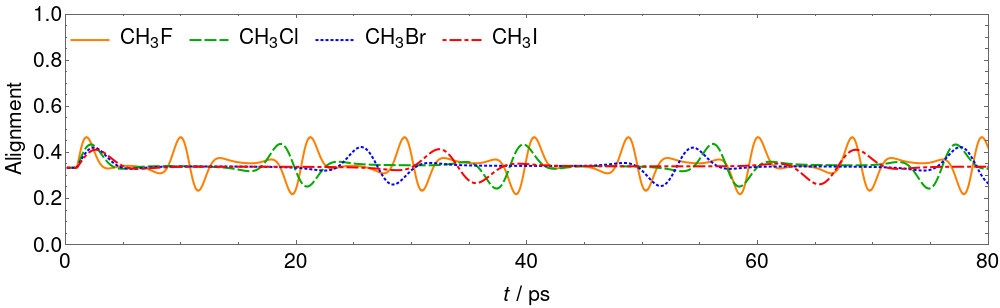}
            \caption{\small{Testing the effect of rotational constants on 
            laser-induced rotational alignment. 
            For all molecules, the polarizability is fixed to values corresponding to CH$_3$Cl.
            The pulse parameters chosen are as follows:
            $T$/K = 5, $\lambda$/nm = 800, $I$/TW\,cm$^{-2}$ = 
            30 (upper panel) and 5 (lower panel), FWHM/fs = 100, and the pulse position is 1 ps.}}
            \label{fgr:CH3X_800nm_pol}
\end{figure*}

\subsubsection{The role of polarizability anisotropy}
Next, let us investigate the relative effects of the rotational constant and the polarizability.
To pinpoint the role of the polarizability anisotropy on laser-induced rotational
alignment, simulations were performed for all molecules, whereby the rotational 
constants were fixed to the value corresponding to CH$_3$F. 
The computed alignment dynamics, shown in Fig.~\ref{fgr:CH3X_800nm_rotconst}, 
confirm that the periodicity of the revivals is determined by the rotational constants,
while the maximal alignment is governed by the polarizability. 
In the lower panel of Fig.~\ref{fgr:CH3X_800nm_rotconst}, 
corresponding to the low-intensity-pulse scenario, the CH$_3$F curve stands out 
with its very small oscillation amplitude, but the peak heights for the 
different molecules reflect the relative values of their polarizability anisotropy. 
In the case of the high-intensity pulse (upper panel of Fig.~\ref{fgr:CH3X_800nm_rotconst}),
CH$_3$F stands out even more with a considerably lower baseline and 
reduced rotational excitation, similar to Fig.~\ref{fgr:CH3X_800nm_alignment}.
The other three molecules show very similar curves in the high-intensity scenario, 
indicating that the differences in their polarizability has a smaller effect 
on the alignment at high pulse intensity than at low intensity, 
due to the saturation of the excitation.

\subsubsection{The role of rotational constants}
The effect the rotational constants have on \AnO\ dynamics was studied by simulations
in which the polarizabilities were fixed to the values corresponding to CH$_3$Cl 
for all molecules.
The results, shown in Fig.~\ref{fgr:CH3X_800nm_pol}, 
demonstrate that the rotational constants not only determine the rotational revival times,
but also affect the alignment maxima, which slightly decrease with the size of the halogen atom.
This behavior can be explained by the increase in the number of rotational states
having significant Boltzmann populations as the size of the halogen atom is increased 
(\textit{i.e.}, the smallest rotational constant is decreased), 
see Table \ref{table:effect_of_T} below. 
The differences in the maximal alignments are more pronounced for the lower intensity pulse,
which indicates a saturation effect in the alignment with respect to the 
laser-pulse intensity, as detailed below in Sec.~\ref{800nmpulseenergy}.
Based on the results obtained by either fixing the rotational constant 
or the polarizability to the same value for all molecules, 
we can conclude that these two parameters have an opposite effect
to the maximal alignment with respect to the size of the halogen atom, 
but the effect of the polarizability dominates, especially at high intensities.

\subsubsection{Sensitivity to the accuracy of molecular parameters}
\begin{figure*}[t!]
            \centering
            \includegraphics[width=\linewidth]{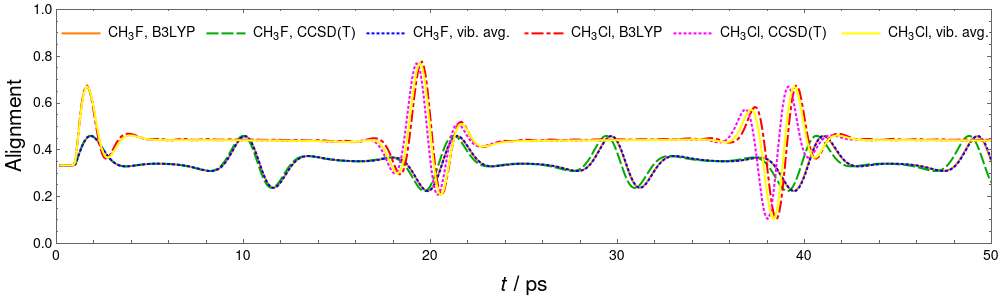}
            \caption{\small{Testing the effect of parameter accuracy on
            the laser-induced rotational alignment of CH$_3$F and CH$_3$Cl. 
            The `B3LYP' and `CCSD(T)' curves correspond to using equilibrium 
            molecular parameters obtained at the
            aug-cc-pV5Z B3LYP and aug-cc-pwCVQZ CCSD(T) levels, respectively
            (see Tables \ref{table:FPA_for_structure} and \ref{table:FPA_for_mu_and_alpha}),
            while the `vib. avg.' curves, which coincide with the `B3LYP' curves,
            were obtained using rotational constants $A_0$ and $B_0$
            (see Table \ref{table:molpropsummary}).}
            The pulse parameters chosen are as follows:
            $T$/K = 5, $\lambda$/nm = 800, $I$/TW\,cm$^{-2}$ = 30,
            FWHM/fs = 100, and the pulse position is 1 ps.}
            \label{fgr:CH3X_800nm_par}
\end{figure*}

In this section we present our results on the effects of molecular parameter accuracy on the laser-induced rotational alignment.
Tables \ref{table:FPA_for_structure} and \ref{table:FPA_for_mu_and_alpha} 
contain the equilibrium molecular parameters of CH$_3$F and CH$_3$Cl computed 
at different levels of electronic-structure theory utilizing various 
basis sets, while Table \ref{table:molpropsummary} contains the 
$A_0$ and $B_0$ parameters.
Figure~\ref{fgr:CH3X_800nm_par} shows the laser-induced rotational alignment 
of CH$_3$F and CH$_3$Cl obtained with parameters computed at the best, 
aug-cc-pwCVQZ CCSD(T), level and at the much less expensive aug-cc-pV5Z B3LYP level. 
The alignment dynamics obtained with the $A_0$ and $B_0$ rotational constants
almost coincides with the related B3LYP curve.
This is due to the fact that $B_0$ is very close to $B_{\rm e,\,a5Z \,\,B3LYP}$.
This agreement is only a coincidence, 
as $A_{\rm e,\,a5Z \,\,B3LYP}$ is closer to $A_{\rm e,\,awcQZ \,\,CCSD(T)}$ than to $A_0$.
The maximal alignments obtained with the different parameter sets 
are very similar to each other but the revivals are drifting away with increasing time.
This indicates that it is crucial to use very accurate rotational constants
in LIMAO simulations.
The accuracy of the polarizability tensor is much less important, 
because it has only a minor effect on the alignment and 
the intensity is not known very precisely in the experiments.

\begin{figure*}[t!]
            \centering
            \includegraphics[width=\linewidth]{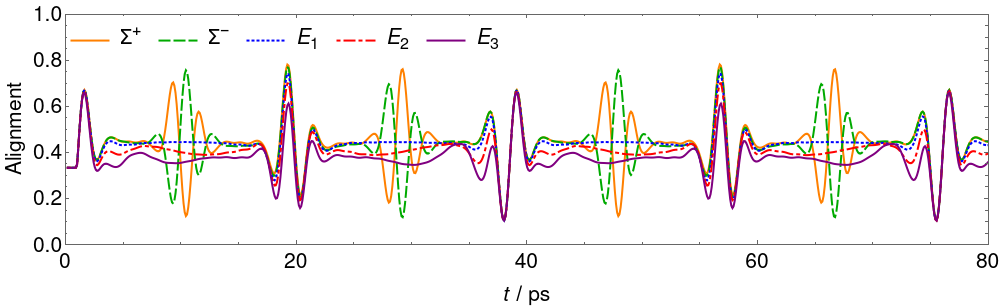}
            \caption{\small{Testing the contribution of the rotational states belonging to each irrep 
            of the $D_\infty$ rotational symmetry group to the alignment of CH$_3$Cl.
            This is based on the selection rules for the optical pulse, 
            which allow  transitions only within a given irrep. 
            The pulse parameters chosen are as follows:
            $T$/K = 5, $\lambda$/nm = 800, $I$/TW\,cm$^{-2}$ = 30,
            FWHM/fs = 100, and the pulse position is 1 ps.}}
            \label{fgr:CH3X_800nm_irrep}
            \end{figure*}

\subsubsection{The role of symmetry and nuclear spin statistical weights}

The selection rules of the laser-induced rotational transitions are determined
by symmetry \cite{06BuJe}. 
The polarizability of the CH$_3$X species transforms as $\Sigma^+$, the totally
symmetric irreducible representation (irrep) of the $D_\infty$ rotation group.
The polarizability transition between states $i$ and $f$, 
belonging to the $\Gamma_i$ and $\Gamma_f$ irreps, respectively, is allowed if 
$\Sigma^+\subseteq \Gamma _f \otimes \Sigma^+ \otimes \Gamma_i=\Gamma _f \otimes \Gamma_i$.
Thus, based on the product table of the $D_\infty$ group \cite{66Herzberg}, 
the polarizability transitions are only allowed between states belonging 
to the same irrep.

In order to study the contribution of the rotational states belonging to each irrep
to the alignment of CH$_3$Cl,
we made simulations where the NSSW was 1 for a single irrep and 0 for the others.
The results can be seen in Fig. \ref{fgr:CH3X_800nm_irrep}.
Rotational states belonging to the $E_1$, $E_2$, $E_3$ irreps 
produce revivals at every $nT_{\rm rev}$ ($n\in \mathbb{N}$), 
while $\Sigma^+$ and $\Sigma^-$ has revivals at every $(n+1/2)T_{\rm rev}$, as well, 
but these cancel each other out when all irreps are included in the simulation.

\begin{figure}[t!]
        \centering
        \includegraphics[width=0.49\linewidth]{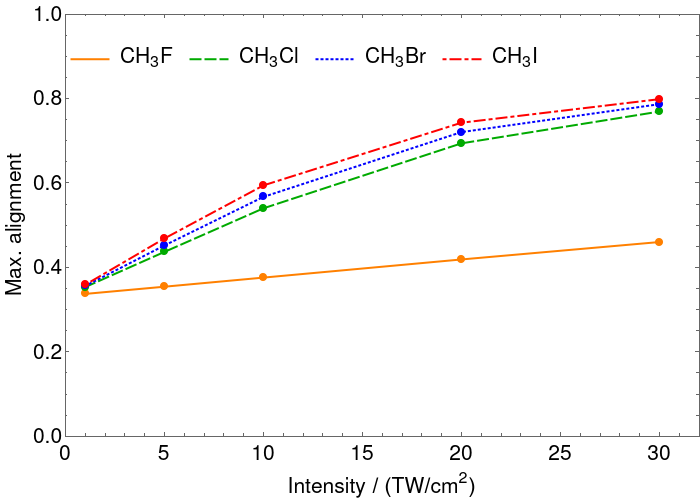}
        \includegraphics[width=0.49\linewidth]{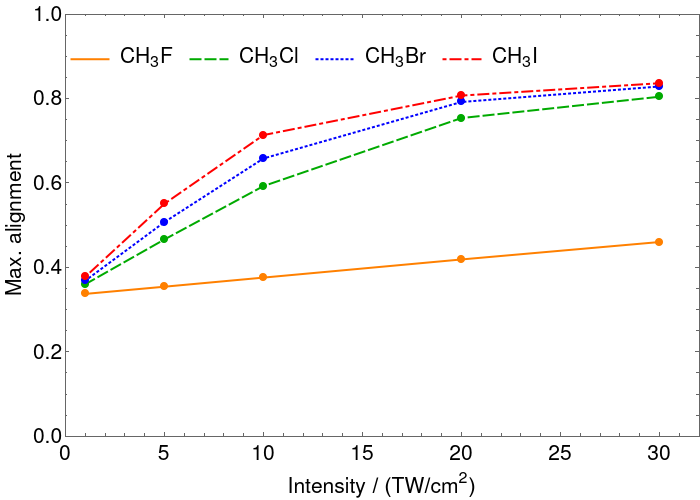}
        \caption{\small{Left panel: 
        The maximal alignment as a function of pulse intensity for the CH$_3$X molecules at 5~K.
        Right panel: same as left panel, but the rotational constants are fixed to the CH$_3$F value for all molecules.  
        The pulse parameters chosen are as follows:
        $T$/K = 5, $\lambda$/nm = 800, FWHM/fs = 100, and the pulse position is 1 ps.}}
        \label{fgr:CH3X_800nm_intensity}
\end{figure}
        
\begin{figure}[t!]
        \centering
        \includegraphics[width=0.49\linewidth]{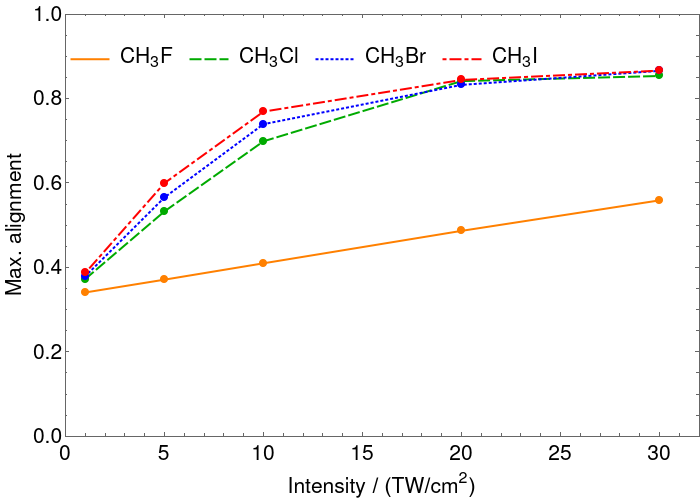}
        \includegraphics[width=0.49\linewidth]{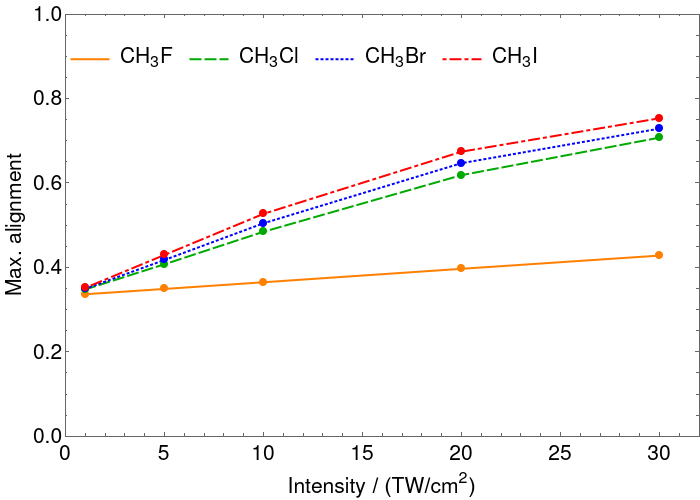}
        \caption{\small{
        The maximal alignment as a function of pulse intensity for the CH$_3$X molecules at 1~K (left panel) and 10~K (right panel).
        The pulse parameters chosen are as follows:
        $\lambda$/nm = 800, FWHM/fs = 100, and the pulse position is 1 ps.}}
        \label{fgr:CH3X_800nm_intensity_10K}
\end{figure}

\subsection{The role of experimental parameters}
    
\subsubsection{\label{800nmpulseenergy}Pulse energy}
    
For the sake of completeness, 
we briefly discuss the effect of experimental parameters on the alignment 
of CH$_3$X molecules.
Figure~\ref{fgr:CH3X_800nm_intensity} shows the maximal alignment as a function 
of the pulse intensity between 1 and 30 TW\,cm$^{-2}$ at 5~K.
The CH$_3$F molecule 
clearly stands out (as also apparent from the previous sections),
because it has a linear response in this intensity range, 
while the behavior of the other three molecules are nonlinear, 
similar in shape, and produce a saturation-type curve. 

The question is whether this saturation-type behavior is a consequence 
of the different Boltzmann populations or the different polarizabilities.
Repeating the above calculation while fixing the rotational constants 
of all molecules to that of the CH$_3$F value 
(see the right panel of Fig.~\ref{fgr:CH3X_800nm_intensity}) 
resulted in a roughly similar plot, which again stresses that CH$_3$F stands out 
because of its much smaller polarizability anisotropy.

\subsubsection{Temperature}
As the atomic weight of the halogen atom and the $r_{\rm C-X}$ carbon-halogen
bond length increase from F to I,
the moments of inertia along the axes perpendicular to the C$-$X bond increase, as well.
This results in the decrease of the respective rotational constants, 
leading to an increased density of rotational states. 
Therefore, at finite temperatures the number of rotational states with 
non-negligible Boltzmann population will increase in succession from F to I,
which is reflected in the values of the rotational partition functions.

Indeed, as summarized in Table~\ref{tbl: field free Q(T) amd population of J},
the value of $Q_{\rm rot}(T)$ increases and the population of the $J=0$ state decreases
when the atomic weight of X (from F to I) increases, \textit{i.e.}, 
the heavier the substituent, the more care is needed to account for all populated 
rotational states. 
With more rotational states having non-negligible Boltzmann population, 
a more pronounced effect of thermal averaging on the \AnO\ dynamics 
should be expected, resulting in lower maximal alignment. 
Figure~\ref{fgr:CH3X_800nm_intensity_10K} shows the maximal alignment 
as a function of the pulse intensity between 1 and 30 TW\,cm$^{-2}$ 
at 1~K and 10~K.
The 10~K case 
exhibits somewhat lower alignment values than those of Fig.~\ref{fgr:CH3X_800nm_intensity},
corresponding to 5 K, while for 1~K, the alignment is higher
and the saturation starts at lower intensities.

\begin{table}[t!]
        \small
        \caption{Rotational partition function ($Q_{\rm rot}(T)$) and population
        of the $J=0$ ground state ($p(J=0)$) at temperatures of 0 K, 0.1 K, and 5 K}
        \label{tbl: field free Q(T) amd population of J}
        \centering
        \begin{tabularx}{\textwidth}{YYYY|YYY}
        \hline \hline
        molecule & \multicolumn{3}{c|}{$Q_{\rm rot}(T)$} & \multicolumn{3}{c}{$p(J=0)$}\\
        
        & $T=0$ K & $T=0.1$ K & $T=5$ K & $T=0$ K & $T=0.1$ K & $T=5$ K \\ \hline
        CH$_3$F  & 2  & 2    & 11.18811 & 1  & 1    & 0.179  \\
        CH$_3$Cl & 2  & 2.00003 & 20.76103 & 1  & 1     & 0.096 \\
        CH$_3$Br & 2  & 2.00076 & 28.31901 & 1  & 1    & 0.071 \\
        CH$_3$I  & 2  & 2.00491 & 32.62516 & 1  & 0.998    & 0.061 \\ \hline\hline
        \end{tabularx}
        \label{table:effect_of_T}
\end{table}

\section{Single-THz-pulse alignment and orientation}
In this section the rotational dynamics of the CH$_3$X species is investigated
when the molecules are excited by a single intense THz pulse, 
causing non-adiabatic \AnO\ dynamics.
Similar to the case of the optical pulse, we study the effect of molecular parameters
and the sensitivity of \AnO\ to the accuracy of these parameters. 
The effects of the pulse parameters are investigated, as well, 
during which the regime of adiabatic \AnO\ dynamics is also explored.

\subsection{The role of pulse parameters, adiabatic and non-adiabatic regimes 
}
\begin{figure*}[t!]
            \centering
            \includegraphics[width=\linewidth]{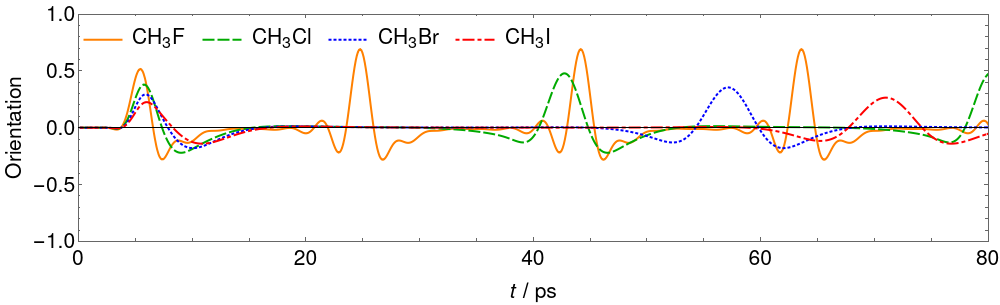} \\
            \includegraphics[width=\linewidth]{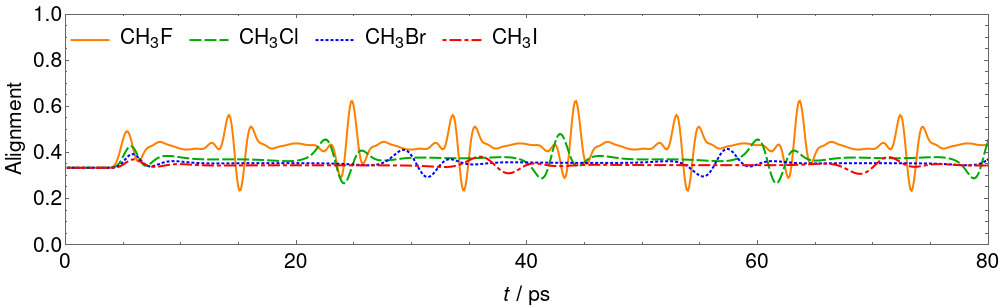}
            \caption{\small {Alignment and orientation dynamics of the CH$_3$X molecules 
            following a THz-pulse excitation with pulse parameters of
            $T$/K = 5,  $\nu$/THz = 0.25, FWHM/fs = 2000, 
            $I$/TW\,cm$^{-2} = 5\times10^{-4}$, pulse position at 5 ps, and CEP = $\pi/2$.}}
            \label{fgr:CH3X_THz_align_orient_1}
\end{figure*}

\begin{figure*}[t!]
            \centering
            \includegraphics[width=\linewidth]{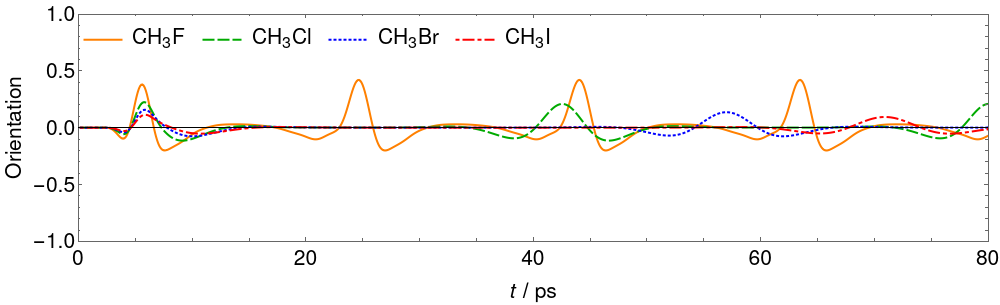} \\
            \includegraphics[width=\linewidth]{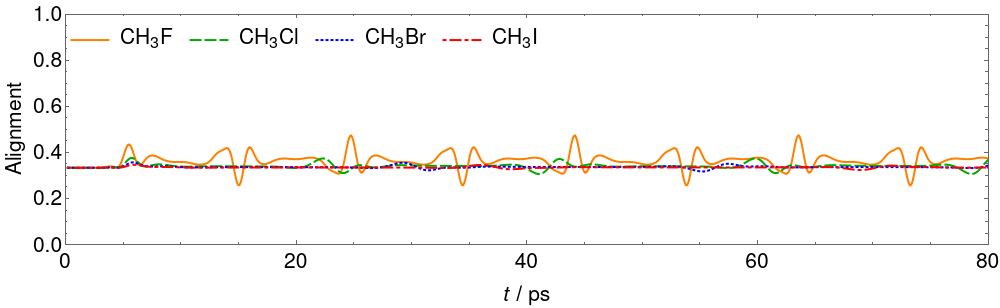}
            \caption{\small {Alignment and orientation dynamics of the CH$_3$X molecules 
            following a THz-pulse excitation with pulse parameters of
            $T$/K = 5,  $\nu$/THz = 0.333, FWHM/fs = 2000, $I$/TW\,cm$^{-2} = 5\times10^{-4}$,
            pulse position at 5 ps, and CEP = $\pi/2$.}}
            \label{fgr:CH3X_THz_align_orient_2}
\end{figure*}

The first set of pulse parameters used in our simulations reflect realistic values 
available at the ELI-ALPS institute \cite{17KuDuKaMo}: 
$\nu=0.333$ THz (equivalent to 
11.1 cm$^{-1}$ wavenumber) or 0.25 THz
(equivalent to 
8.3 cm$^{-1}$ wavenumber),
$I=5\times10^{-4}$ TW\,cm$^{-2}$, FWHM $=2000$ fs, and  CEP $=\pi/2$, 
where CEP stands for carrier envelope phase. 
In the case of THz pulses, the field-dipole interaction becomes dominant, 
because the polarizability interaction is proportional to the square of the electric field
and the intensity is much lower than it was for the optical pulses. 
Furthermore, contrary to the optical-pulse case, 
the electric-field oscillation is slow enough for the field-dipole interaction
not to average out to zero (cycle averaging can not be used), 
which can lead to orientation in addition to alignment.
As presented in Figs.~\ref{fgr:CH3X_THz_align_orient_1} and \ref{fgr:CH3X_THz_align_orient_2},
such THz pulses can produce significant alignment and orientation for CH$_3$F,
with smaller effect for the other molecules. 
The order of maximal orientation and alignment is CH$_3$F $>$ CH$_3$Cl $>$ CH$_3$Br $>$ CH$_3$I. 
In accordance with Section \ref{Theory}, 
the alignment revivals are alternating and have the same time period as in the single-optical pulse case, while the orientation revivals do not alternate and occur twice as seldom.

The maximal alignment and orientation is larger for the 0.25 THz pulse
than for the 0.333 THz pulse, because the photon energy of the former 
is closer to the rotational transitions determined by the 
$B_{\rm e}$ rotational constants (see Table~\ref{table:molpropsummary}).

\begin{figure*}[t!]
            \centering
            \includegraphics[width=\linewidth]{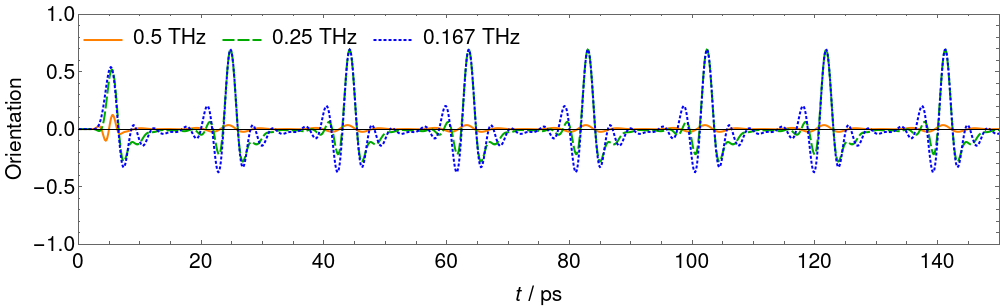} \\
            \includegraphics[width=\linewidth]{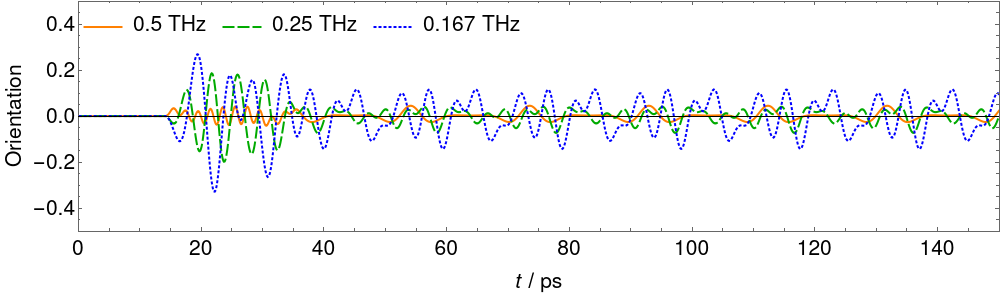} \\
            \includegraphics[width=\linewidth]{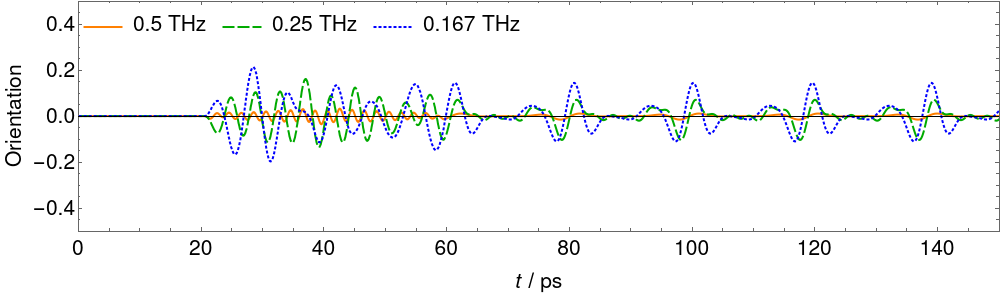} \\
            \caption{\small {Effect of THz pulse parameters on the orientation of CH$_3$F.
            The frequency ($\nu$), intensity ($I$), and FWHM was varied, keeping the total pulse energy, $I\times$FWHM constant. 
            $T$/K = 5  and CEP = $\pi/2$ were used for all panels. The other pulse parameters of the top panel:
            FWHM=2 ps, $I=5\times10^{-4}$ TW\,cm$^{-2}$,
            pulse position was 5 ps; 
            middle panel: FWHM=10 ps, $I=1\times10^{-4}$ TW\,cm$^{-2}$, pulse position was 25 ps;
            bottom panel: FWHM=20 ps, $I=5\times10^{-5}$ TW\,cm$^{-2}$, pulse position was 40 ps.}}
            \label{fgr:CH3X_THz_pulse_param}
\end{figure*}

\begin{figure*}[t!]
            \centering
            \includegraphics[width=\linewidth]{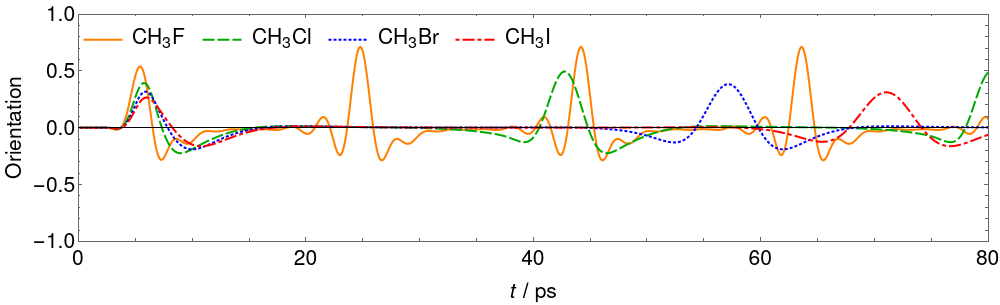}
            \\
            \includegraphics[width=\linewidth]{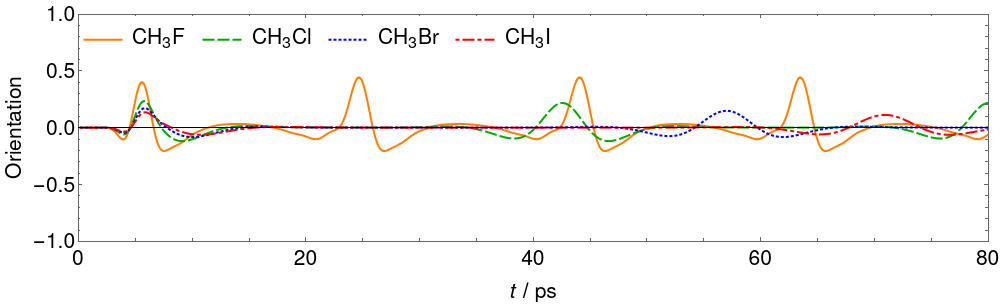}
            \caption{Testing the effect of the molecular parameters in 
            the alignment and orientation of the CH$_3$X molecules for a $\nu$/THz = 0.25 (upper panel)
            and a 0.333 (lower panel) pulse. 
            The polarizability was set to zero and the dipole moment was 2~D for all molecules.
            Pulse parameters: $T$/K = 5, FWHM/fs = 2000, pulse position is 5 ps, 
            $I$/TW\,cm$^{-2} = 5\times10^{-4}$,
            CEP = $\pi/2$.}
            \label{fgr:CH3X_THz_alpha_0_mu_2D}
\end{figure*}


The \textit{adiabatic} regime of the \AnO\ dynamics was also studied
by varying the THz pulse parameters.
We chose CH$_{3}$F for this investigation,
because the other molecules would provide much smaller adiabatic \AnO.
\textit{Non-adiabatic alignment or orientation} 
is observed when the laser pulse is turned on (and/or off) much
faster than the rotational period(s). In this case
the molecule shows time-dependent revivals under field-free conditions,
long after the laser pulse disappeared.
In the case of \textit{adiabatic alignment or orientation}, 
the laser pulse is turned on (and off) much slower than
the rotational period(s). In this case
the spatial order of the molecules exists only during the pulse. 
Figure \ref{fgr:CH3X_THz_pulse_param} shows how the orientation for CH$_3$F
is affected by the frequency, duration, and intensity of the pulse.
The intensity and the 
duration of the pulse were varied such that $I\times$FWHM
was kept constant. 
The tested frequencies are 0.5, 0.25, and 0.167 THz, the intensity values are 
$5\times10^{-4}$, $1\times10^{-4}$, and $5\times10^{-5}$  TW\,cm$^{-2}$,
while 2, 10, and 20 ps are used for the FWHM.
One can see that the FWHM = 2 ps case is clearly non-adiabatic, 
because the pulse is much shorter than the characteristic timescale of rotational motion,
but for the cases of FWHM = 10 and 20 ps, we are entering the adiabatic regime, 
though the maximal orientation significantly decreases 
(note that the range of the vertical axis is only [$-0.5$, 0.5]).
In this case, the molecular orientation follows the oscillation of the 
electric field during the pulse, but  revivals are still present after the pulse,
so the situation is not completely adiabatic. 
Out of the curves shown in Fig. \ref{fgr:CH3X_THz_pulse_param},
the $\nu=0.25$ THz and FWHM = 10 ps case is the closest to the adiabatic situation,
but the achieved orientation is only moderate. 
The maximal orientation decreases if the frequency is increased, 
because the frequency is moved further from the rotational transitions,
so the excitation decreases. 
Fig. \ref{fgr:CH3X_THz_pulse_param} demonstrates that for a given frequency, increasing the pulse duration also decreases 
the orientation and the degree of excitation, because the bandwidth 
of the pulse becomes narrower, and less resonant with the rotational transitions.

\subsection{The role of molecular parameters}
    
\subsubsection{Rotational constants, dipole moment, and polarizability}
    
The computations summarized in Figs. \ref{fgr:CH3X_THz_align_orient_1} 
and \ref{fgr:CH3X_THz_align_orient_2} were repeated after setting the polarizabilities
to zero for all four molecules, which did not change the results within numerical error.
Therefore, the polarizability interaction with the field is negligible in this situation,
as expected at such low field intensity.
To clarify the different THz pulse-induced rotational behavior of the CH$_3$X species 
and the role of the rotational constants,
simulations with zero polarizability and an equal dipole of  2 D
for all molecules were carried out. 
(in fact, the dipole moment is quite similar for all molecules, between 1.7--1.9 D).
As can be seen in Fig.~\ref{fgr:CH3X_THz_alpha_0_mu_2D}, 
changes in the orientation with respect to changing the halogen atom are nearly the same as in 
Figs.~\ref{fgr:CH3X_THz_align_orient_1} and \ref{fgr:CH3X_THz_align_orient_2};
therefore, it is understood that the rotational constants are responsible 
for the different molecular behavior.
If the size of the halogen is increased, more rotational states will be populated
at a given finite temperature, which decreases the maximal orientation. 
Furthermore, increasing the size of the halogen decreases the transition frequency
between the rotational states, so the transition frequencies are further away from 
the central frequency of the pulse, which decreases the degree of excitation.     
    
The computations summarized in 
Figs. \ref{fgr:CH3X_THz_align_orient_1} and \ref{fgr:CH3X_THz_align_orient_2}
were repeated after setting the rotational constant to the CH$_3$F value for all molecules, 
resulting in nearly identical orientation for all the molecules.
This confirms that the rotational constants determine the orientation in this situation, 
\textit{i.e.}, when the molecules have similar permanent dipole moments.

\subsubsection{Sensitivity to the accuracy of molecular parameters}
\begin{figure*}[t!]
            \centering
            \includegraphics[width=\linewidth]{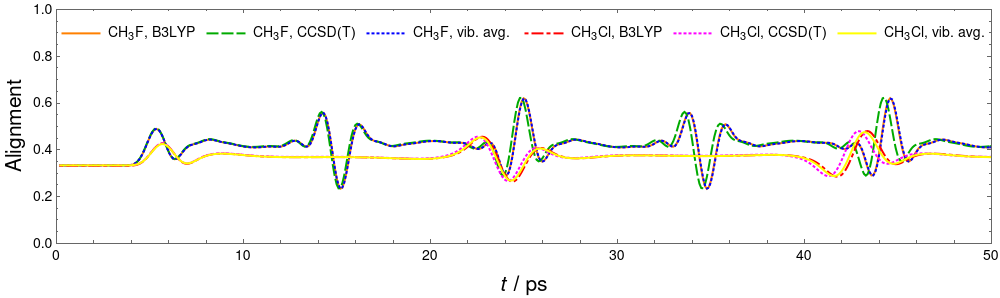}
            \\
            \includegraphics[width=\linewidth]{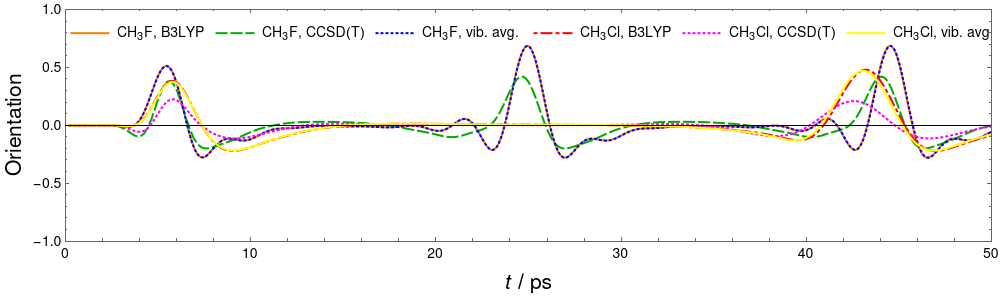}
            \caption{\small {Testing the effect of parameter accuracy on
            the alignment and orientation of CH$_3$F and CH$_3$Cl. 
            The upper and lower panels show the alignment and orientation, respectively.
            The `B3LYP' and `CCSD(T)' curves correspond to using equilibrium molecular parameters obtained at the
            aug-cc-pV5Z B3LYP and aug-cc-pwCVQZ CCSD(T) levels
            (see Tables \ref{table:FPA_for_structure} and \ref{table:FPA_for_mu_and_alpha}),
            while the `vib. avg.' curve (which accidentally coincides with the `B3LYP' curve) was obtained using and $A_0$ and $B_0$ rotational constants, and $\mu_0$ dipole moment (see Table \ref{table:molpropsummary}). }
            The pulse parameters chosen are as follows: $T$/K = 5, $\nu$/THz = 0.25, FWHM/fs = 2000,
            pulse position at 5 ps, $I$/TW\,cm$^{-2} = 5\times10^{-4}$, and CEP = $\pi/2$.}
            \label{fgr:CH3X_THz_par}
\end{figure*}

Figure \ref{fgr:CH3X_THz_par} shows the \AnO\ dynamics of CH$_3$F and CH$_3$Cl,
obtained with parameters computed either at the highest, aug-cc-pwCVQZ CCSD(T) level 
or the much less expensive aug-cc-pV5Z B3LYP level,
and also with using $A_0$, $B_0$, and $\mu_0$ instead of the equilibrium values.
Similarly to the case of the optical pulse, 
the results obtained with the vibrationally averaged parameters 
and with the aug-cc-pV5Z B3LYP equilibrium parameters coincide.

One can observe from Fig.~\ref{fgr:CH3X_THz_par} that the revivals computed 
with the different parameter sets are not only drifting away with increasing time,
but the maximal orientation is very different. 
This further emphasizes the need to use accurate rotational constants 
to obtain accurate orientation information.

\subsubsection{
The role of symmetry and nuclear-spin-statistical weights}
\begin{figure*}[t!]
            \centering
            \includegraphics[width=\linewidth]{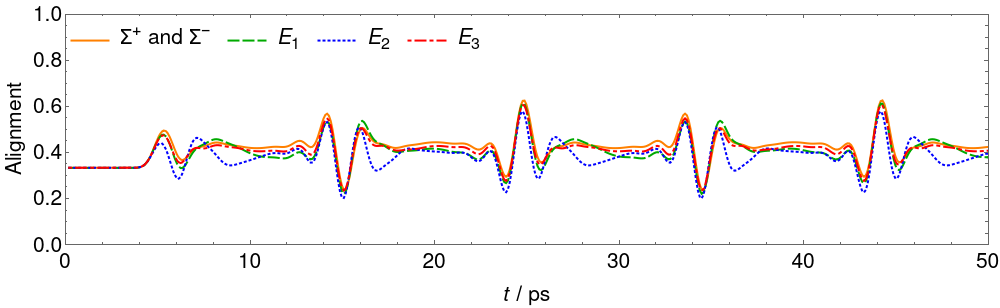}
            \\
            \includegraphics[width=\linewidth]{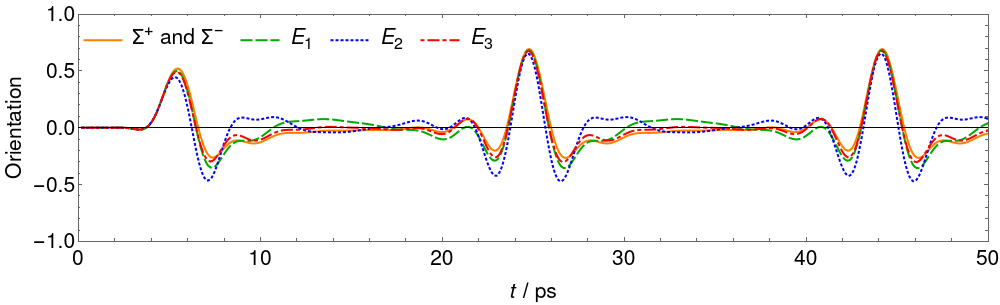}
            \caption{\small{
            Role of symmetry in the \AnO\ dynamics of CH$_3$F (see text for details).
            Pulse parameters: $T$/K = 5, $\nu$/THz = 0.25, FWHM/fs = 2000, 
            pulse position is 5 ps, $I$/TW\,cm$^{-2} = 5\times10^{-4}$, and
            CEP = $\pi/2$.}}
            \label{fgr:CH3X_THz_irrep}
\end{figure*}

To complete the picture, we analyzed the role of symmetry for the THz pulse, as well. The dipole moment of CH$_3$X transforms as the $\Sigma^-$ irrep of the 
$D_\infty$ rotational group. 
This determines the dipole transition selection rules: 
the transition between states $i$ and $f$, belonging to the $\Gamma_i$ and $\Gamma_f$ irreps,
respectively, is allowed if $\Sigma^+\subseteq \Gamma _f \otimes \Sigma^- \otimes \Gamma_i$, 
\textit{i.e}, the direct product contains the total symmetric irrep. 
Thus, the allowed transitions are the following:
$\Sigma^+\leftrightarrow\Sigma^-$, $E_1\leftrightarrow E_1$, $E_2\leftrightarrow E_2$, and 
$E_3\leftrightarrow E_3$.
In order to study how transitions with different symmetry contribute to the \AnO\ dynamics 
of CH$_3$F, we made separate simulations for the subsets of irreps interconnected
by the dipole interaction (see Fig. \ref{fgr:CH3X_THz_irrep}), 
where the NSSW of the given subset was 1, and all the others 0. 
These subsets are $\{\Sigma^+, \Sigma^-\}$, $\{E_1\}$, $\{E_2\}$, $\{E_3\}$.
The \AnO\ curves are quite similar in each case, except for slight differences in the $\{E_2\}$ curve towards the end of each revival. Contrary to the 
optical pulse scenario (see Fig. \ref{fgr:CH3X_800nm_irrep}), 
no new revivals appear in Fig. \ref{fgr:CH3X_THz_irrep} with respect to Fig. \ref{fgr:CH3X_THz_align_orient_1}.

\section{Summary and conclusions}
The laser-induced alignment and orientation (\AnO) dynamics of the prolate
symmetric-top CH$_3$X (X = F, Cl, Br and I) molecules were investigated using 
standard quantum-chemistry software and the in-house LIMAO \cite{18SzJoYa} code. 
The molecular parameters required by LIMAO for the \AnO\ simulations, 
\textit{i.e.}, the rotational constants, the dipole moment vectors, 
and the polarizability tensors, 
were computed using different levels of 
electronic-structure theory and employing a variety of basis sets.
The numerical convergence of the molecular parameters toward the ultimate 
complete-basis-set full-configuration-interaction limit was investigated 
utilizing the focal-point-analysis (FPA) approach \cite{93AlEaCs,98CsAlSc}.
The molecular parameters computed 
and employed 
in the LIMAO simulations correspond to the equilibrium structure.
Table \ref{table:molpropsummary} of this paper summarizes our 
most accurate estimates of these values.
For CH$_3$F and CH$_3$Cl, vibrationally averaged rotational constants
($A_0$ and $B_0$) and dipoles ($\mu_0$) were also determined and 
utilized in LIMAO simulations.

The qualitative changes observed in the molecular parameters computed 
can be explained straightforwardly and these trends can be used 
to understand the effect of halogen substitution on the laser-induced 
\AnO\ dynamics. 
The $B_{\rm e}$ rotational constant 
decreases in the F$\to$ Cl$\to$ Br$\to$ I order,
that is as the halogen gets heavier.
Changes in $A_{\rm e}$ upon halogen substitution is much less pronounced and not monotonic.
The dipole moments vary considerably less than the other molecular properties
and it is a non-monotonous function of the atomic number of the halogen, 
having a maximum for CH$_3$Cl. 
Qualitatively, the dipole is determined by the partial charges on the 
carbon and halogen atoms, which depend on the electronegativity
of these atoms, and the charge separation,
which depends on the carbon-halogen bond length.
These two properties show opposite trends with the increasing size of the halogen, 
and their effect on the dipole moment tend to cancel each other out.
The polarizability anisotropy, $\Delta\alpha$, 
which determines the polarizability (field-induced dipole) interaction, 
increases with the size of the halogen: 
$\Delta \alpha_{\rm CH_3F}<\Delta \alpha_{\rm CH_3Cl}<\Delta \alpha_{\rm CH_3Br}<\Delta \alpha_{\rm CH_3I}$, with the largest difference found between X $=$ F and Cl.

Sensitivity of the \AnO\ results on the molecular parameters was
tested extensively, showing that the \AnO\ dynamics are most sensitive 
to the accuracy of the rotational constants, especially that of $B$. 
Interestingly, the \AnO\ curves obtained using vibrationally averaged
rotational constants and dipole moments are very similar to those resulting 
from the use of the equilibrium values of these parameters computed at 
the inexpensive aug-cc-pV5Z B3LYP level.
Using equilibrium molecular parameters obtained at the most 
sophisticated, aug-cc-pwCVQZ CCSD(T) level, results in slightly different
\AnO\ dynamics.
This can be explained by the fact that, coincidentally,
$B_0$ and $B_{\rm e,\,a5Z\,\,B3LYP}$
happen to be very close to each other, while $B_{\rm e,\,awcQZ\,\,CCSD(T)}$
is a bit different from both of them.
This relationship does not characterize the $A$ rotational constants.

Our systematic computations demonstrate that
a short and intense optical laser pulse,
with a chosen wavelength of 800 nm, can be used to achieve non-adiabatic alignment, 
whereby the molecule interacts with the electric field through its polarizability.
The time between the alternating alignment revivals (half revivals), 
$T_{\rm rev}=1/(4cB_{\rm e})$,
is inversely proportional to the rotational constant;
therefore, it increases with the size of the halogen. 
The maximal alignment is much smaller for CH$_3$F than for the other molecules.
The maximal alignment for CH$_3$F depends linearly on the pulse intensity, 
at least in the interval studied, while the other three molecules show 
a saturation-type behavior. 
The maximal alignment decreases if the rotational constant is decreased,
but increases with increased polarizability anisotropy. 
For the CH$_3$X molecules, the latter effect dominates, 
especially at large pulse intensities, resulting in much greater alignment 
for CH$_3$Cl, CH$_3$Br, and CH$_3$I than for CH$_3$F.
This behavior reflects the 
polarizability anisotropy of the four molecules.
The contribution of the rotational states with different symmetry, 
weighed by  their nuclear spin statistical weights (NSSW), was determined, 
demonstrating 
that certain alignment revivals cancel each other out in the averaged signal. 
Increasing the temperature decreases the maximal alignment, as expected.

Short and intense THz laser pulses can cause non-adiabatic orientation
(and alignment) in the sample, with the time between the orientation
revivals being twice as long as for the alignment. 
The field-dipole interaction is the dominant one in the case of THz pulses, 
and the interaction due to the polarizability of the systems
can be neglected. 
The frequency of the THz pulses is in the range of the rotational transitions;
however, for few-cycle pulses the central frequency does not have to be resonant
with a specific transition for achieving \AnO. 
Due to the short duration of the pulse, 
the bandwidth covers several rotational transitions.
We investigated the orientation of the CH$_3$X molecules induced 
by experimentally feasible, intense few-cycle THz pulses with $\nu=0.25$ THz 
(equivalent to 8.3 \cm) and 0.333 THz (equivalent to 11.1 \cm) frequency 
and having a full width at half maximum 
of 2 picoseconds. 
For these pulses, the maximal orientation decreases with the size of the halogen;
thus, contrary to the case of the optical pulse, 
CH$_3$F has the largest maximal \AnO\ among the four molecules.
The rotational constants determine the time between the revivals, 
similar to the optical-pulse case, but it is again the rotational constant
which mainly determines the maximal orientation, 
as the CH$_3$X molecules have similar dipole moments. 
If the rotational constants are decreased, i.e., the size of the 
halogen is increased, more states will be populated 
at a given finite temperature, which lowers the maximal orientation. 
The transition frequency of the rotational transitions also decreases with 
decreasing rotational constants, leading to a smaller overlap between the 
bandwidth of the laser pulse and the rotational transitions 
(note that the rotational transitions of the CH$_3$X molecules relevant 
in the \AnO\ dynamics simulations of this study are mostly smaller than the 
THz photon energies). 
In the case of CH$_3$F, we also studied the (nearly) adiabatic regime
of the THz-pulse-induced \AnO, showing much lower \AnO\ than in the 
non-adiabatic case.
A detailed investigation of the relation between the laser-induced rotational populations and the resulting \AnO\ dynamics is planned in a future work.


\section{Acknowledgements}

This research was supported by the 
J\'{a}nos Bolyai Research Scholarship
of the Hungarian Academy of Sciences, awarded to TSz,
and by the \'{U}NKP-20-5 and \'{U}NKP-20-3 
New National Excellence Program of the Ministry for Innovation and Technology 
from the source of the National Research, Development and Innovation Fund,
awarded to TSz and IS, respectively.
The Hungarian co-authors are also grateful to NKFIH for additional support 
(Grants No. FK134291, K119658, and K138233).
ELI-ALPS is supported by the European Union and co-financed by the European Regional Development Fund (GI-NOP-2.3.6-15-2015-00001). KC and MUK acknowledge Project no. 2019-2.1.13-TÉT-IN-2020-00059 which has been implemented with the support provided from the National Research, Development and Innovation Fund of Hungary, financed under the 2019-2.1.13-TÉT-IN funding scheme. 

\section{Data availability}
The data that support the findings of this study are available from the corresponding author upon reasonable request.

\section{Conflict of interest}
The authors have no conflicts to disclose.


\bibliographystyle{mod_elsarticle-num-names}
\bibliography{master,journals,paperone}

\begin{thebibliography}{78}
\expandafter\ifx\csname natexlab\endcsname\relax\def\natexlab#1{#1}\fi
\providecommand{\url}[1]{\texttt{#1}}
\providecommand{\href}[2]{#2}
\providecommand{\path}[1]{#1}
\providecommand{\DOIprefix}{doi:}
\providecommand{\ArXivprefix}{arXiv:}
\providecommand{\URLprefix}{URL: }
\providecommand{\Pubmedprefix}{pmid:}
\providecommand{\doi}[1]{\href{http://dx.doi.org/#1}{\path{#1}}}
\providecommand{\Pubmed}[1]{\href{pmid:#1}{\path{#1}}}
\providecommand{\bibinfo}[2]{#2}
\ifx\xfnm\relax \def\xfnm[#1]{\unskip,\space#1}\fi
\bibitem[{Stapelfeldt and Seideman(2003)}]{03StSe}
\bibinfo{author}{H.~Stapelfeldt}, \bibinfo{author}{T.~Seideman},
\newblock \bibinfo{title}{Colloquium: Aligning molecules with strong laser
  pulses},
\newblock \bibinfo{journal}{Rev. Mod. Phys.} \bibinfo{volume}{75}
  (\bibinfo{year}{2003}) \bibinfo{pages}{543--557}.
\bibitem[{Ohshima and Hasegawa(2010)}]{10OhHa}
\bibinfo{author}{Y.~Ohshima}, \bibinfo{author}{H.~Hasegawa},
\newblock \bibinfo{title}{Coherent rotational excitation by intense nonresonant
  laser fields},
\newblock \bibinfo{journal}{Int. Rev. Phys. Chem.} \bibinfo{volume}{29}
  (\bibinfo{year}{2010}) \bibinfo{pages}{619--663}.
\bibitem[{Lemeshko et~al.(2013)Lemeshko, Krems, Doyle, and Kais}]{13LeKrDoKa}
\bibinfo{author}{M.~Lemeshko}, \bibinfo{author}{R.~V. Krems},
  \bibinfo{author}{J.~M. Doyle}, \bibinfo{author}{S.~Kais},
\newblock \bibinfo{title}{Manipulation of molecules with electromagnetic
  fields},
\newblock \bibinfo{journal}{Mol. Phys.} \bibinfo{volume}{111}
  (\bibinfo{year}{2013}) \bibinfo{pages}{1648--1682}.
\bibitem[{Koch et~al.(2019)Koch, Lemeshko, and Sugny}]{19KoLeSu}
\bibinfo{author}{C.~P. Koch}, \bibinfo{author}{M.~Lemeshko},
  \bibinfo{author}{D.~Sugny},
\newblock \bibinfo{title}{Quantum control of molecular rotation},
\newblock \bibinfo{journal}{Rev. Mod. Phys.} \bibinfo{volume}{91}
  (\bibinfo{year}{2019}) \bibinfo{pages}{035005}.
\bibitem[{Härtelt and Friedrich(2008)}]{08MaBr}
\bibinfo{author}{M.~Härtelt}, \bibinfo{author}{B.~Friedrich},
\newblock \bibinfo{title}{Directional states of symmetric-top molecules
  produced by combined static and radiative electric fields},
\newblock \bibinfo{journal}{J. Chem. Phys.} \bibinfo{volume}{128}
  (\bibinfo{year}{2008}) \bibinfo{pages}{224313}.
\bibitem[{Korobenko and Milner(2016)}]{16KoMi}
\bibinfo{author}{A.~Korobenko}, \bibinfo{author}{V.~Milner},
\newblock \bibinfo{title}{Adiabatic field-free alignment of asymmetric top
  molecules with an optical centrifuge},
\newblock \bibinfo{journal}{Phys. Rev. Lett.} \bibinfo{volume}{116}
  (\bibinfo{year}{2016}) \bibinfo{pages}{183001}.
\bibitem[{Larsen et~al.(1999)Larsen, Sakai, Safvan, Wendt-Larsen, and
  Stapelfeldt}]{99LaSaSaWe}
\bibinfo{author}{J.~J. Larsen}, \bibinfo{author}{H.~Sakai},
  \bibinfo{author}{C.~P. Safvan}, \bibinfo{author}{I.~Wendt-Larsen},
  \bibinfo{author}{H.~Stapelfeldt},
\newblock \bibinfo{title}{Aligning molecules with intense nonresonant laser
  fields},
\newblock \bibinfo{journal}{J. Chem. Phys.} \bibinfo{volume}{111}
  (\bibinfo{year}{1999}) \bibinfo{pages}{7774--7781.}
\bibitem[{Friedrich and Herschbach(1995)}]{95FrHe}
\bibinfo{author}{B.~Friedrich}, \bibinfo{author}{D.~Herschbach},
\newblock \bibinfo{title}{Alignment and trapping of molecules in intense laser
  fields},
\newblock \bibinfo{journal}{Phys. Rev. Lett.} \bibinfo{volume}{74}
  (\bibinfo{year}{1995}) \bibinfo{pages}{4623--4626.}
\bibitem[{Bisgaard et~al.(2004)Bisgaard, Poulsen, P{\'{e}}ronne, Viftrup, and
  Stapelfeldt}]{04BiPoPeVi}
\bibinfo{author}{C.~Z. Bisgaard}, \bibinfo{author}{M.~D. Poulsen},
  \bibinfo{author}{E.~P{\'{e}}ronne}, \bibinfo{author}{S.~S. Viftrup},
  \bibinfo{author}{H.~Stapelfeldt},
\newblock \bibinfo{title}{Observation of enhanced field-free molecular
  alignment by two laser pulses},
\newblock \bibinfo{journal}{Phys. Rev. Lett.} \bibinfo{volume}{92}
  (\bibinfo{year}{2004}) \bibinfo{pages}{173004}.
\bibitem[{Damari et~al.(2016)Damari, Kallush, and Fleischer}]{16DaKaFl}
\bibinfo{author}{R.~Damari}, \bibinfo{author}{S.~Kallush},
  \bibinfo{author}{S.~Fleischer},
\newblock \bibinfo{title}{Rotational control of asymmetric molecules: Dipole-
  versus polarizability-driven rotational dynamics},
\newblock \bibinfo{journal}{Phys. Rev. Lett.} \bibinfo{volume}{117}
  (\bibinfo{year}{2016}) \bibinfo{pages}{103001}.
\bibitem[{Sonoda et~al.(2018)Sonoda, Iwasaki, Yamanouchi, and
  Hasegawa}]{18SoIwYaHa}
\bibinfo{author}{K.~Sonoda}, \bibinfo{author}{A.~Iwasaki},
  \bibinfo{author}{K.~Yamanouchi}, \bibinfo{author}{H.~Hasegawa},
\newblock \bibinfo{title}{Field-free molecular orientation of nonadiabatically
  aligned {OCS}},
\newblock \bibinfo{journal}{Chem. Phys. Lett.} \bibinfo{volume}{693}
  (\bibinfo{year}{2018}) \bibinfo{pages}{114--120}.
\bibitem[{Rosca-Pruna and Vrakking(2001)}]{01RoVr}
\bibinfo{author}{F.~Rosca-Pruna}, \bibinfo{author}{M.~Vrakking},
\newblock \bibinfo{title}{{Experimental Observation of Revival Structures in
  Picosecond Laser-Induced Alignment of I$_2$}},
\newblock \bibinfo{journal}{Phys. Rev. Lett.} \bibinfo{volume}{87}
  (\bibinfo{year}{2001}) \bibinfo{pages}{153902}.
\bibitem[{Seideman(1999)}]{99Se}
\bibinfo{author}{T.~Seideman},
\newblock \bibinfo{title}{Revival structure of aligned rotational wave
  packets},
\newblock \bibinfo{journal}{Phys. Rev. Lett.} \bibinfo{volume}{83}
  (\bibinfo{year}{1999}) \bibinfo{pages}{4971--4974.}
\bibitem[{Shepperson et~al.(2017)Shepperson, Chatterley, S{\o}ndergaard,
  Christiansen, Lemeshko, and Stapelfeldt}]{17ShChSoCh}
\bibinfo{author}{B.~Shepperson}, \bibinfo{author}{A.~S. Chatterley},
  \bibinfo{author}{A.~A. S{\o}ndergaard}, \bibinfo{author}{L.~Christiansen},
  \bibinfo{author}{M.~Lemeshko}, \bibinfo{author}{H.~Stapelfeldt},
\newblock \bibinfo{title}{Strongly aligned molecules inside helium droplets in
  the near-adiabatic regime},
\newblock \bibinfo{journal}{J. Chem. Phys.} \bibinfo{volume}{147}
  (\bibinfo{year}{2017}) \bibinfo{pages}{013946}.
\bibitem[{Pickering et~al.(2019)Pickering, Shepperson, Christiansen, and
  Stapelfeldt}]{19PiShChSt}
\bibinfo{author}{J.~D. Pickering}, \bibinfo{author}{B.~Shepperson},
  \bibinfo{author}{L.~Christiansen}, \bibinfo{author}{H.~Stapelfeldt},
\newblock \bibinfo{title}{Alignment of the {${\mathrm{CS}}_{2}$} dimer embedded
  in helium droplets induced by a circularly polarized laser pulse},
\newblock \bibinfo{journal}{Phys. Rev. A} \bibinfo{volume}{99}
  (\bibinfo{year}{2019}) \bibinfo{pages}{043403}.
\bibitem[{Lee et~al.(2006)Lee, Villeneuve, Corkum, Stolow, and
  Underwood}]{06LeViCoSt}
\bibinfo{author}{K.~F. Lee}, \bibinfo{author}{D.~M. Villeneuve},
  \bibinfo{author}{P.~B. Corkum}, \bibinfo{author}{A.~Stolow},
  \bibinfo{author}{J.~G. Underwood},
\newblock \bibinfo{title}{Field-free three-dimensional alignment of polyatomic
  molecules},
\newblock \bibinfo{journal}{Phys. Rev. Lett.} \bibinfo{volume}{97}
  (\bibinfo{year}{2006}) \bibinfo{pages}{173001}.
\bibitem[{Artamonov and Seideman(2012)}]{12ArSe}
\bibinfo{author}{M.~Artamonov}, \bibinfo{author}{T.~Seideman},
\newblock \bibinfo{title}{Three-dimensional laser alignment of polyatomic
  molecular ensembles},
\newblock \bibinfo{journal}{Mol. Phys.} \bibinfo{volume}{110}
  (\bibinfo{year}{2012}) \bibinfo{pages}{885--896}.
\bibitem[{Ren et~al.(2014)Ren, Makhija, and Kumarappan}]{14ReMaKu}
\bibinfo{author}{X.~Ren}, \bibinfo{author}{V.~Makhija},
  \bibinfo{author}{V.~Kumarappan},
\newblock \bibinfo{title}{Multipulse three-dimensional alignment of asymmetric
  top molecules},
\newblock \bibinfo{journal}{Phys. Rev. Lett.} \bibinfo{volume}{112}
  (\bibinfo{year}{2014}) \bibinfo{pages}{173602}.
\bibitem[{Larsen et~al.(2000)Larsen, Hald, Bjerre, Stapelfeldt, and
  Seideman}]{00LaHaBjSt}
\bibinfo{author}{J.~J. Larsen}, \bibinfo{author}{K.~Hald},
  \bibinfo{author}{N.~Bjerre}, \bibinfo{author}{H.~Stapelfeldt},
  \bibinfo{author}{T.~Seideman},
\newblock \bibinfo{title}{Three dimensional alignment of molecules using
  elliptically polarized laser fields},
\newblock \bibinfo{journal}{Phys. Rev. Lett.} \bibinfo{volume}{85}
  (\bibinfo{year}{2000}) \bibinfo{pages}{2470--2473.}
\bibitem[{Tutunnikov et~al.(2021)Tutunnikov, Xu, Field, Nelson, Prior, and
  Averbukh}]{21TuXuFiNe}
\bibinfo{author}{I.~Tutunnikov}, \bibinfo{author}{L.~Xu},
  \bibinfo{author}{R.~W. Field}, \bibinfo{author}{K.~A. Nelson},
  \bibinfo{author}{Y.~Prior}, \bibinfo{author}{I.~S. Averbukh},
\newblock \bibinfo{title}{Enantioselective orientation of chiral molecules
  induced by terahertz pulses with twisted polarization},
\newblock \bibinfo{journal}{Phys. Rev. Res.} \bibinfo{volume}{3}
  (\bibinfo{year}{2021}) \bibinfo{pages}{013249}.
\bibitem[{Fleischer et~al.(2011)Fleischer, Zhou, Field, and Nelson}]{11ZhFiNe}
\bibinfo{author}{S.~Fleischer}, \bibinfo{author}{Y.~Zhou},
  \bibinfo{author}{R.~W. Field}, \bibinfo{author}{K.~A. Nelson},
\newblock \bibinfo{title}{Molecular orientation and alignment by intense
  single-cycle {THz} pulses},
\newblock \bibinfo{journal}{Phys. Rev. Lett.} \bibinfo{volume}{107}
  (\bibinfo{year}{2011}) \bibinfo{pages}{163603}.
\bibitem[{Felker(1992)}]{92Felker}
\bibinfo{author}{P.~M. Felker},
\newblock \bibinfo{title}{Rotational coherence spectroscopy: studies of the
  geometries of large gas-phase species by picosecond time-domain methods},
\newblock \bibinfo{journal}{J. Phys. Chem.} \bibinfo{volume}{96}
  (\bibinfo{year}{1992}) \bibinfo{pages}{7844--7857}.
\bibitem[{Schr\"{o}ter et~al.(2018)Schr\"{o}ter, Lee, and Schultz}]{18ScLeSc}
\bibinfo{author}{C.~Schr\"{o}ter}, \bibinfo{author}{J.~C. Lee},
  \bibinfo{author}{T.~Schultz},
\newblock \bibinfo{title}{Mass-correlated rotational raman spectra with high
  resolution, broad bandwidth, and absolute frequency accuracy},
\newblock \bibinfo{journal}{Proc. Natl. Acad. Sci.} \bibinfo{volume}{115}
  (\bibinfo{year}{2018}) \bibinfo{pages}{5072--5076,}.
\bibitem[{Riehn(2002)}]{02Riehn}
\bibinfo{author}{C.~Riehn},
\newblock \bibinfo{title}{High-resolution pump{\textendash}probe rotational
  coherence spectroscopy {\textendash} rotational constants and structure of
  ground and electronically excited states of large molecular systems},
\newblock \bibinfo{journal}{Chem. Phys.} \bibinfo{volume}{283}
  (\bibinfo{year}{2002}) \bibinfo{pages}{297--329.}
\bibitem[{Chatterley et~al.(2020)Chatterley, Christiansen, Schouder,
  J{\o}rgensen, Shepperson, Cherepanov, Bighin, Zillich, Lemeshko, and
  Stapelfeldt}]{20ChChScJo}
\bibinfo{author}{A.~S. Chatterley}, \bibinfo{author}{L.~Christiansen},
  \bibinfo{author}{C.~A. Schouder}, \bibinfo{author}{A.~V. J{\o}rgensen},
  \bibinfo{author}{B.~Shepperson}, \bibinfo{author}{I.~N. Cherepanov},
  \bibinfo{author}{G.~Bighin}, \bibinfo{author}{R.~E. Zillich},
  \bibinfo{author}{M.~Lemeshko}, \bibinfo{author}{H.~Stapelfeldt},
\newblock \bibinfo{title}{Rotational coherence spectroscopy of molecules in
  helium nanodroplets: Reconciling the time and the frequency domains},
\newblock \bibinfo{journal}{Phys. Rev. Lett.} \bibinfo{volume}{125}
  (\bibinfo{year}{2020}) \bibinfo{pages}{013001}.
\bibitem[{Nisoli et~al.(2017)Nisoli, Decleva, Calegari, Palacios, and
  Mart{\'{\i}}n}]{17NiDeCaPa}
\bibinfo{author}{M.~Nisoli}, \bibinfo{author}{P.~Decleva},
  \bibinfo{author}{F.~Calegari}, \bibinfo{author}{A.~Palacios},
  \bibinfo{author}{F.~Mart{\'{\i}}n},
\newblock \bibinfo{title}{Attosecond electron dynamics in molecules},
\newblock \bibinfo{journal}{Chem. Rev.} \bibinfo{volume}{117}
  (\bibinfo{year}{2017}) \bibinfo{pages}{10760--10825}.
\bibitem[{Krausz and Ivanov(2009)}]{09KrIv}
\bibinfo{author}{F.~Krausz}, \bibinfo{author}{M.~Ivanov},
\newblock \bibinfo{title}{Attosecond physics},
\newblock \bibinfo{journal}{Rev. Mod. Phys.} \bibinfo{volume}{81}
  (\bibinfo{year}{2009}) \bibinfo{pages}{163--234}.
\bibitem[{K\"{u}hn et~al.(2017)K\"{u}hn, Dumergue, Kahaly, Mondal, F\"{u}le,
  Csizmadia, Farkas, Major, V{\'{a}}rallyay, Cormier, Kalashnikov, Calegari,
  Devetta, Frassetto, M{\aa}nsson, Poletto, Stagira, Vozzi, Nisoli, Rudawski,
  Maclot, Campi, Wikmark, Arnold, Heyl, Johnsson, L'Huillier, Lopez-Martens,
  Haessler, Bocoum, Boehle, Vernier, Iaquaniello, Skantzakis, Papadakis,
  Kalpouzos, Tzallas, L{\'{e}}pine, Charalambidis, Varj{\'{u}}, Osvay, and
  Sansone}]{17KuDuKaMo}
\bibinfo{author}{S.~K\"{u}hn}, \bibinfo{author}{M.~Dumergue},
  \bibinfo{author}{S.~Kahaly}, \bibinfo{author}{S.~Mondal},
  \bibinfo{author}{M.~F\"{u}le}, \bibinfo{author}{T.~Csizmadia},
  \bibinfo{author}{B.~Farkas}, \bibinfo{author}{B.~Major},
  \bibinfo{author}{Z.~V{\'{a}}rallyay}, \bibinfo{author}{E.~Cormier},
  \bibinfo{author}{M.~Kalashnikov}, \bibinfo{author}{F.~Calegari},
  \bibinfo{author}{M.~Devetta}, \bibinfo{author}{F.~Frassetto},
  \bibinfo{author}{E.~M{\aa}nsson}, \bibinfo{author}{L.~Poletto},
  \bibinfo{author}{S.~Stagira}, \bibinfo{author}{C.~Vozzi},
  \bibinfo{author}{M.~Nisoli}, \bibinfo{author}{P.~Rudawski},
  \bibinfo{author}{S.~Maclot}, \bibinfo{author}{F.~Campi},
  \bibinfo{author}{H.~Wikmark}, \bibinfo{author}{C.~L. Arnold},
  \bibinfo{author}{C.~M. Heyl}, \bibinfo{author}{P.~Johnsson},
  \bibinfo{author}{A.~L'Huillier}, \bibinfo{author}{R.~Lopez-Martens},
  \bibinfo{author}{S.~Haessler}, \bibinfo{author}{M.~Bocoum},
  \bibinfo{author}{F.~Boehle}, \bibinfo{author}{A.~Vernier},
  \bibinfo{author}{G.~Iaquaniello}, \bibinfo{author}{E.~Skantzakis},
  \bibinfo{author}{N.~Papadakis}, \bibinfo{author}{C.~Kalpouzos},
  \bibinfo{author}{P.~Tzallas}, \bibinfo{author}{F.~L{\'{e}}pine},
  \bibinfo{author}{D.~Charalambidis}, \bibinfo{author}{K.~Varj{\'{u}}},
  \bibinfo{author}{K.~Osvay}, \bibinfo{author}{G.~Sansone},
\newblock \bibinfo{title}{The {ELI}-{ALPS} facility: the next generation of
  attosecond sources},
\newblock \bibinfo{journal}{J. Phys. B} \bibinfo{volume}{50}
  (\bibinfo{year}{2017}) \bibinfo{pages}{132002}.
\bibitem[{Kraus et~al.(2015)Kraus, Mignolet, Baykusheva, Rupenyan, Horn{\'{y}},
  Penka, Grassi, Tolstikhin, Schneider, Jensen, Madsen, Bandrauk, Remacle, and
  W\"{o}rner}]{15KrMiBaRu}
\bibinfo{author}{P.~M. Kraus}, \bibinfo{author}{B.~Mignolet},
  \bibinfo{author}{D.~Baykusheva}, \bibinfo{author}{A.~Rupenyan},
  \bibinfo{author}{L.~Horn{\'{y}}}, \bibinfo{author}{E.~F. Penka},
  \bibinfo{author}{G.~Grassi}, \bibinfo{author}{O.~I. Tolstikhin},
  \bibinfo{author}{J.~Schneider}, \bibinfo{author}{F.~Jensen},
  \bibinfo{author}{L.~B. Madsen}, \bibinfo{author}{A.~D. Bandrauk},
  \bibinfo{author}{F.~Remacle}, \bibinfo{author}{H.~J. W\"{o}rner},
\newblock \bibinfo{title}{Measurement and laser control of attosecond charge
  migration in ionized iodoacetylene},
\newblock \bibinfo{journal}{Science} \bibinfo{volume}{350}
  (\bibinfo{year}{2015}) \bibinfo{pages}{790--795}.
\bibitem[{Bisgaard et~al.(2009)Bisgaard, Clarkin, Wu, Lee, Gessner, Hayden, and
  Stolow}]{09BiClWuLe}
\bibinfo{author}{C.~Z. Bisgaard}, \bibinfo{author}{O.~J. Clarkin},
  \bibinfo{author}{G.~Wu}, \bibinfo{author}{A.~M.~D. Lee},
  \bibinfo{author}{O.~Gessner}, \bibinfo{author}{C.~C. Hayden},
  \bibinfo{author}{A.~Stolow},
\newblock \bibinfo{title}{Time-resolved molecular frame dynamics of
  fixed-in-space {CS}$_2$ molecules},
\newblock \bibinfo{journal}{Science} \bibinfo{volume}{323}
  (\bibinfo{year}{2009}) \bibinfo{pages}{1464--1468.}
\bibitem[{Madsen et~al.(2009)Madsen, Madsen, Viftrup, Johansson, Poulsen,
  Holmegaard, Kumarappan, J{\o}rgensen, and Stapelfeldt}]{09MaMaViJo}
\bibinfo{author}{C.~B. Madsen}, \bibinfo{author}{L.~B. Madsen},
  \bibinfo{author}{S.~S. Viftrup}, \bibinfo{author}{M.~P. Johansson},
  \bibinfo{author}{T.~B. Poulsen}, \bibinfo{author}{L.~Holmegaard},
  \bibinfo{author}{V.~Kumarappan}, \bibinfo{author}{K.~A. J{\o}rgensen},
  \bibinfo{author}{H.~Stapelfeldt},
\newblock \bibinfo{title}{Manipulating the torsion of molecules by strong laser
  pulses},
\newblock \bibinfo{journal}{Phys. Rev. Lett.} \bibinfo{volume}{102}
  (\bibinfo{year}{2009}) \bibinfo{pages}{073007}.
\bibitem[{Skantzakis et~al.(2016)Skantzakis, Chatziathanasiou, Carpeggiani,
  Sansone, Nayak, Gray, Tzallas, Charalambidis, Hertz, and
  Faucher}]{16SkChCaSa}
\bibinfo{author}{E.~Skantzakis}, \bibinfo{author}{S.~Chatziathanasiou},
  \bibinfo{author}{P.~A. Carpeggiani}, \bibinfo{author}{G.~Sansone},
  \bibinfo{author}{A.~Nayak}, \bibinfo{author}{D.~Gray},
  \bibinfo{author}{P.~Tzallas}, \bibinfo{author}{D.~Charalambidis},
  \bibinfo{author}{E.~Hertz}, \bibinfo{author}{O.~Faucher},
\newblock \bibinfo{title}{Polarization shaping of high-order harmonics in
  laser-aligned molecules},
\newblock \bibinfo{journal}{Sci. Rep.} \bibinfo{volume}{6}
  (\bibinfo{year}{2016}) \bibinfo{pages}{39295}.
\bibitem[{Faucher et~al.(2016)Faucher, Prost, Hertz, Billard, Lavorel, Milner,
  Milner, Zyss, and Averbukh}]{16FaPrHeBi}
\bibinfo{author}{O.~Faucher}, \bibinfo{author}{E.~Prost},
  \bibinfo{author}{E.~Hertz}, \bibinfo{author}{F.~Billard},
  \bibinfo{author}{B.~Lavorel}, \bibinfo{author}{A.~A. Milner},
  \bibinfo{author}{V.~A. Milner}, \bibinfo{author}{J.~Zyss},
  \bibinfo{author}{I.~S. Averbukh},
\newblock \bibinfo{title}{Rotational {Doppler} effect in harmonic generation
  from spinning molecules},
\newblock \bibinfo{journal}{Phys. Rev. A} \bibinfo{volume}{94}
  (\bibinfo{year}{2016}) \bibinfo{pages}{051402}.
\bibitem[{Yang et~al.(2021)Yang, Liu, Zhao, Tu, Liu, and Zhao}]{21YaLiZhTu}
\bibinfo{author}{Y.~Yang}, \bibinfo{author}{L.~Liu}, \bibinfo{author}{J.~Zhao},
  \bibinfo{author}{Y.~Tu}, \bibinfo{author}{J.~Liu}, \bibinfo{author}{Z.~Zhao},
\newblock \bibinfo{title}{Effect of ionization asymmetry on high harmonic
  generation from oriented {CO} in orthogonal two-color fields},
\newblock \bibinfo{journal}{J. Phys. B-At. Mol. Opt.} \bibinfo{volume}{54}
  (\bibinfo{year}{2021}) \bibinfo{pages}{144009}.
\bibitem[{Hamilton et~al.(2005)Hamilton, Seideman, Ejdrup, Poulsen, Bisgaard,
  Viftrup, and Stapelfeldt}]{05HaSeEjPo}
\bibinfo{author}{E.~Hamilton}, \bibinfo{author}{T.~Seideman},
  \bibinfo{author}{T.~Ejdrup}, \bibinfo{author}{M.~D. Poulsen},
  \bibinfo{author}{C.~Z. Bisgaard}, \bibinfo{author}{S.~S. Viftrup},
  \bibinfo{author}{H.~Stapelfeldt},
\newblock \bibinfo{title}{Alignment of symmetric top molecules by short laser
  pulses},
\newblock \bibinfo{journal}{Phys. Rev. A} \bibinfo{volume}{72}
  (\bibinfo{year}{2005}) \bibinfo{pages}{043402}.
\bibitem[{Luo et~al.(2017)Luo, Hu, Yu, Zhu, He, Li, Ma, Wang, Liu, Roeterdink,
  Stolte, and Ding}]{17LuHuYuZh}
\bibinfo{author}{S.~Luo}, \bibinfo{author}{W.~Hu}, \bibinfo{author}{J.~Yu},
  \bibinfo{author}{R.~Zhu}, \bibinfo{author}{L.~He}, \bibinfo{author}{X.~Li},
  \bibinfo{author}{P.~Ma}, \bibinfo{author}{C.~Wang}, \bibinfo{author}{F.~Liu},
  \bibinfo{author}{W.~G. Roeterdink}, \bibinfo{author}{S.~Stolte},
  \bibinfo{author}{D.~Ding},
\newblock \bibinfo{title}{Rotational dynamics of quantum state-selected
  symmetric-top molecules in nonresonant femtosecond laser fields},
\newblock \bibinfo{journal}{J. Phys. Chem. A} \bibinfo{volume}{121}
  (\bibinfo{year}{2017}) \bibinfo{pages}{777--783}.
\bibitem[{Luo et~al.(2018)Luo, Zhou, Hu, Yu, Li, Ma, He, Wang, Guo, Yang, and
  Ding}]{18LuZhHuYu}
\bibinfo{author}{S.~Luo}, \bibinfo{author}{S.~Zhou}, \bibinfo{author}{W.~Hu},
  \bibinfo{author}{J.~Yu}, \bibinfo{author}{X.~Li}, \bibinfo{author}{P.~Ma},
  \bibinfo{author}{L.~He}, \bibinfo{author}{C.~Wang}, \bibinfo{author}{F.~Guo},
  \bibinfo{author}{Y.~Yang}, \bibinfo{author}{D.~Ding},
\newblock \bibinfo{title}{Identifying the multielectron effect on chemical bond
  rearrangement of {CH$_3$Cl} molecules in strong laser fields},
\newblock \bibinfo{journal}{J. Phys. Chem. A} \bibinfo{volume}{122}
  (\bibinfo{year}{2018}) \bibinfo{pages}{8427--8432}.
\bibitem[{He et~al.(2016)He, Pan, Yang, Luo, Lu, Zhao, Li, Song, Stolte, Ding,
  and Roeterdink}]{16HePaYaLu}
\bibinfo{author}{L.~He}, \bibinfo{author}{Y.~Pan}, \bibinfo{author}{Y.~Yang},
  \bibinfo{author}{S.~Luo}, \bibinfo{author}{C.~Lu}, \bibinfo{author}{H.~Zhao},
  \bibinfo{author}{D.~Li}, \bibinfo{author}{L.~Song},
  \bibinfo{author}{S.~Stolte}, \bibinfo{author}{D.~Ding},
  \bibinfo{author}{W.~G. Roeterdink},
\newblock \bibinfo{title}{Ion yields of laser aligned {CH$_3$I} and {CH$_3$Br}
  from multiple orbitals},
\newblock \bibinfo{journal}{Chem. Phys. Lett.} \bibinfo{volume}{665}
  (\bibinfo{year}{2016}) \bibinfo{pages}{141--146}.
\bibitem[{S{\'{a}}ndor et~al.(2019)S{\'{a}}ndor, Sissay, Mauger, Gordon,
  Gorman, Scarborough, Gaarde, Lopata, Schafer, and Jones}]{19SaSiMaGo}
\bibinfo{author}{P.~S{\'{a}}ndor}, \bibinfo{author}{A.~Sissay},
  \bibinfo{author}{F.~Mauger}, \bibinfo{author}{M.~W. Gordon},
  \bibinfo{author}{T.~T. Gorman}, \bibinfo{author}{T.~D. Scarborough},
  \bibinfo{author}{M.~B. Gaarde}, \bibinfo{author}{K.~Lopata},
  \bibinfo{author}{K.~J. Schafer}, \bibinfo{author}{R.~R. Jones},
\newblock \bibinfo{title}{Angle-dependent strong-field ionization of
  halomethanes},
\newblock \bibinfo{journal}{J. Chem. Phys.} \bibinfo{volume}{151}
  (\bibinfo{year}{2019}) \bibinfo{pages}{194308}.
\bibitem[{Szidarovszky et~al.(2018)Szidarovszky, Jono, and
  Yamanouchi}]{18SzJoYa}
\bibinfo{author}{T.~Szidarovszky}, \bibinfo{author}{M.~Jono},
  \bibinfo{author}{K.~Yamanouchi},
\newblock \bibinfo{title}{{LIMAO}: Cross-platform software for simulating
  laser-induced alignment and orientation dynamics of linear-, symmetric- and
  asymmetric tops},
\newblock \bibinfo{journal}{Comp. Phys. Commun.} \bibinfo{volume}{228}
  (\bibinfo{year}{2018}) \bibinfo{pages}{219--228}.
\bibitem[{Helgaker et~al.(2000)Helgaker, J{\o{}}rgensen, and Olsen}]{00HeJoOl}
\bibinfo{author}{T.~Helgaker}, \bibinfo{author}{P.~J{\o{}}rgensen},
  \bibinfo{author}{J.~Olsen}, \bibinfo{title}{{Molecular Electronic-Structure
  Theory}}, \bibinfo{publisher}{Wiley, New York}, \bibinfo{year}{2000}.
\bibitem[{Hasegawa and Ohshima(2015)}]{15HaOh}
\bibinfo{author}{H.~Hasegawa}, \bibinfo{author}{Y.~Ohshima},
  \bibinfo{title}{Nonadiabatic Molecular Alignment and Orientation},
  \bibinfo{publisher}{Springer International Publishing},
  \bibinfo{address}{Cham}, \bibinfo{year}{2015}, pp. \bibinfo{pages}{45--64}.
\bibitem[{Zare(1988)}]{88Zare}
\bibinfo{author}{R.~N. Zare}, \bibinfo{title}{{Angular Momentum: Understanding
  Spatial Aspects in Chemistry and Physics}}, \bibinfo{publisher}{Wiley, New
  York}, \bibinfo{year}{1988}.
\bibitem[{Bunker and Jensen(2006)}]{06BuJe}
\bibinfo{author}{P.~R. Bunker}, \bibinfo{author}{P.~Jensen},
  \bibinfo{title}{{Molecular Symmetry and Spectroscopy}},
  \bibinfo{publisher}{NRC Research Press, Ottawa}, \bibinfo{year}{2006}.
\bibitem[{\url{ http://www.cfour.de}(????)}]{cfour}
\url{ http://www.cfour.de}, \bibinfo{title}{{CFOUR, a quantum chemical program
  package, last accessed on \today}}.
\bibitem[{\url{ https://www.molpro.net/ }(????)}]{molpro}
\url{ https://www.molpro.net/ }, \bibinfo{title}{{MOLPRO website, last accessed
  on \today}}.
\bibitem[{Hehre et~al.(1986)Hehre, Radom, v.~R.~Schleyer, and
  Pople}]{86HeRaScPo}
\bibinfo{author}{W.~J. Hehre}, \bibinfo{author}{L.~Radom},
  \bibinfo{author}{P.~v.~R.~Schleyer}, \bibinfo{author}{J.~A. Pople},
  \bibinfo{title}{Ab Initio Molecular Orbital Theory},
  \bibinfo{publisher}{Wiley-Interscience, New York}, \bibinfo{year}{1986}.
\bibitem[{Jensen(2006)}]{06Jensen}
\bibinfo{author}{F.~Jensen}, \bibinfo{title}{{Introduction to Computational
  Chemistry}}, \bibinfo{publisher}{Wiley, Chichester}, \bibinfo{year}{2006}.
\bibitem[{Roothaan(1951)}]{51Roothaan}
\bibinfo{author}{C.~C.~J. Roothaan},
\newblock \bibinfo{title}{New developments in molecular orbital theory},
\newblock \bibinfo{journal}{Rev. Mod. Phys.} \bibinfo{volume}{23}
  (\bibinfo{year}{1951}) \bibinfo{pages}{69--89}.
\bibitem[{M\o{}ller and Plesset(1934)}]{34MoPe}
\bibinfo{author}{C.~M\o{}ller}, \bibinfo{author}{M.~S. Plesset},
\newblock \bibinfo{title}{Note on an approximation treatment for many-electron
  systems},
\newblock \bibinfo{journal}{Phys. Rev.} \bibinfo{volume}{46}
  (\bibinfo{year}{1934}) \bibinfo{pages}{618--622}.
\bibitem[{Krishnan et~al.(1980)Krishnan, Frisch, and Pople}]{80KrFrPo}
\bibinfo{author}{R.~Krishnan}, \bibinfo{author}{M.~J. Frisch},
  \bibinfo{author}{J.~A. Pople},
\newblock \bibinfo{title}{Contribution of triple substitutions to the electron
  correlation energy in fourth order perturbation theory},
\newblock \bibinfo{journal}{J. Chem. Phys.} \bibinfo{volume}{72}
  (\bibinfo{year}{1980}) \bibinfo{pages}{4244--4245}.
\bibitem[{Čížek(1966)}]{66Cizek}
\bibinfo{author}{J.~Čížek},
\newblock \bibinfo{title}{On the correlation problem in atomic and molecular
  systems. {C}alculation of wavefunction components in {U}rsell‐type
  expansion using quantum‐field theoretical methods},
\newblock \bibinfo{journal}{J. Chem. Phys.} \bibinfo{volume}{45}
  (\bibinfo{year}{1966}) \bibinfo{pages}{4256--4266}.
\bibitem[{Purvis and Bartlett(1982)}]{82PuBa}
\bibinfo{author}{G.~D. Purvis}, \bibinfo{author}{R.~J. Bartlett},
\newblock \bibinfo{title}{A full coupled‐cluster singles and doubles model:
  The inclusion of disconnected triples},
\newblock \bibinfo{journal}{J. Chem. Phys.} \bibinfo{volume}{76}
  (\bibinfo{year}{1982}) \bibinfo{pages}{1910--1918}.
\bibitem[{Raghavachari et~al.(1989)Raghavachari, Trucks, Pople, and
  Head-Gordon}]{89RaTrPoHe}
\bibinfo{author}{K.~Raghavachari}, \bibinfo{author}{G.~W. Trucks},
  \bibinfo{author}{J.~A. Pople}, \bibinfo{author}{M.~Head-Gordon},
\newblock \bibinfo{title}{{A fifth-order perturbation comparison of electron
  correlation theories}},
\newblock \bibinfo{journal}{Chem. Phys. Lett.} \bibinfo{volume}{157}
  (\bibinfo{year}{1989}) \bibinfo{pages}{479--483}.
\bibitem[{Becke(1988)}]{88Becke}
\bibinfo{author}{A.~D. Becke},
\newblock \bibinfo{title}{Density-functional exchange-energy approximation with
  correct asymptotic behavior},
\newblock \bibinfo{journal}{Phys. Rev. A} \bibinfo{volume}{38}
  (\bibinfo{year}{1988}) \bibinfo{pages}{3098--3100}.
\bibitem[{Becke(1993)}]{93Becke}
\bibinfo{author}{A.~D. Becke},
\newblock \bibinfo{title}{Density‐functional thermochemistry. {III. T}he role
  of exact exchange},
\newblock \bibinfo{journal}{J. Chem. Phys.} \bibinfo{volume}{98}
  (\bibinfo{year}{1993}) \bibinfo{pages}{5648--5652}.
\bibitem[{Lee et~al.(1988)Lee, Yang, and Parr}]{88LeYaPa}
\bibinfo{author}{C.~Lee}, \bibinfo{author}{W.~Yang}, \bibinfo{author}{R.~G.
  Parr},
\newblock \bibinfo{title}{Development of the {Colle--Salvetti}
  correlation-energy formula into a functional of the electron density},
\newblock \bibinfo{journal}{Phys. Rev. B} \bibinfo{volume}{37}
  (\bibinfo{year}{1988}) \bibinfo{pages}{785--789}.
\bibitem[{{Dunning Jr.}(1989)}]{89Dunning}
\bibinfo{author}{T.~H. {Dunning Jr.}},
\newblock \bibinfo{title}{{Gaussian basis sets for use in correlated molecular
  calculations. I. The atoms boron through neon and hydrogen}},
\newblock \bibinfo{journal}{J. Chem. Phys.} \bibinfo{volume}{90}
  (\bibinfo{year}{1989}) \bibinfo{pages}{1007--1023}.
\bibitem[{Allen et~al.(1993)Allen, East, and Cs\'asz\'ar}]{93AlEaCs}
\bibinfo{author}{W.~D. Allen}, \bibinfo{author}{A.~L.~L. East},
  \bibinfo{author}{A.~G. Cs\'asz\'ar},
\newblock \bibinfo{title}{\textit{Ab initio} anharmonic vibrational analyses of
  non-rigid moleculer},
\newblock in: \bibinfo{editor}{J.~Laane}, \bibinfo{editor}{M.~Dakkouri},
  \bibinfo{editor}{B.~van~der Veken}, \bibinfo{editor}{H.~Oberhammer} (Eds.),
  \bibinfo{booktitle}{Structures and conformations of nonrigid molecules},
  \bibinfo{publisher}{Kluwer, Dordrecht}, \bibinfo{year}{1993}, pp.
  \bibinfo{pages}{343--373}.
\bibitem[{Cs\'asz\'ar et~al.(1998)Cs\'asz\'ar, Allen, and {Schaefer
  III}}]{98CsAlSc}
\bibinfo{author}{A.~G. Cs\'asz\'ar}, \bibinfo{author}{W.~D. Allen},
  \bibinfo{author}{H.~F. {Schaefer III}},
\newblock \bibinfo{title}{{In pursuit of the \textit{ab initio} limit for
  conformational energy prototypes}},
\newblock \bibinfo{journal}{J. Chem. Phys.} \bibinfo{volume}{108}
  (\bibinfo{year}{1998}) \bibinfo{pages}{9751--9764}.
\bibitem[{Mantina et~al.(2009)Mantina, Chamberlin, Valero, Cramer, and
  Truhlar}]{mantina2009consistent}
\bibinfo{author}{M.~Mantina}, \bibinfo{author}{A.~C. Chamberlin},
  \bibinfo{author}{R.~Valero}, \bibinfo{author}{C.~J. Cramer},
  \bibinfo{author}{D.~G. Truhlar},
\newblock \bibinfo{title}{Consistent van der {Waals} radii for the whole main
  group},
\newblock \bibinfo{journal}{J. Phys. Chem. A} \bibinfo{volume}{113}
  (\bibinfo{year}{2009}) \bibinfo{pages}{5806--5812}.
\bibitem[{lid(2004)}]{lide2004CRCHandbook}
\bibinfo{title}{CRC Handbook of Chemistry and Physics, 84th Edition Edited by
  David R. Lide (National Institute of Standards and Technology).}, volume
  \bibinfo{volume}{126}, \bibinfo{year}{2004}.
\bibitem[{Schwerdtfeger and Nagle(2019)}]{schwerdtfeger2019table}
\bibinfo{author}{P.~Schwerdtfeger}, \bibinfo{author}{J.~K. Nagle},
\newblock \bibinfo{title}{2018 table of static dipole polarizabilities of the
  neutral elements in the periodic table},
\newblock \bibinfo{journal}{Mol. Phys.} \bibinfo{volume}{117}
  (\bibinfo{year}{2019}) \bibinfo{pages}{1200--1225}.
\bibitem[{Demaison et~al.(1999)Demaison, Breidung, Thiel, and
  Papousek}]{99DeBrThPa}
\bibinfo{author}{J.~Demaison}, \bibinfo{author}{J.~Breidung},
  \bibinfo{author}{W.~Thiel}, \bibinfo{author}{D.~Papousek},
\newblock \bibinfo{title}{The equilibrium structure of methyl fluoride},
\newblock \bibinfo{journal}{Struct. Chem.} \bibinfo{volume}{10}
  (\bibinfo{year}{1999}) \bibinfo{pages}{129--133}.
\bibitem[{Jensen et~al.(1981)Jensen, Brodersen, and Guelachvili}]{81JeBrGu}
\bibinfo{author}{P.~Jensen}, \bibinfo{author}{S.~Brodersen},
  \bibinfo{author}{G.~Guelachvili},
\newblock \bibinfo{title}{Determination of {$A_0$} for {CH$_3^{~35}$Cl and
  CH$_3^{~37}$Cl} from the $\nu_4$ infrared and {R}aman bands},
\newblock \bibinfo{journal}{J. Mol. Spectrosc.} \bibinfo{volume}{88}
  (\bibinfo{year}{1981}) \bibinfo{pages}{378--393}.
\bibitem[{Graner(1981)}]{94Graner}
\bibinfo{author}{G.~Graner},
\newblock \bibinfo{title}{The methyl bromide molecule: {A} critical
  consideration of perturbations in spectra},
\newblock \bibinfo{journal}{J. Mol. Spectrosc.} \bibinfo{volume}{90}
  (\bibinfo{year}{1981}) \bibinfo{pages}{394--438}.
\bibitem[{Matsuura and Overend(1972)}]{72MaOv}
\bibinfo{author}{H.~Matsuura}, \bibinfo{author}{J.~Overend},
\newblock \bibinfo{title}{Equilibrium structure of methyl iodide},
\newblock \bibinfo{journal}{J. Chem. Phys.} \bibinfo{volume}{56}
  (\bibinfo{year}{1972}) \bibinfo{pages}{5725--5727}.
\bibitem[{Papousek et~al.(1993)Papousek, Hsu, Chen, Pracna, Klee, and
  Winnewisser}]{93PaHsChPr}
\bibinfo{author}{D.~Papousek}, \bibinfo{author}{Y.~Hsu},
  \bibinfo{author}{H.~Chen}, \bibinfo{author}{P.~Pracna},
  \bibinfo{author}{S.~Klee}, \bibinfo{author}{M.~Winnewisser},
\newblock \bibinfo{title}{{Far Infrared Spectrum and Ground State Parameters of
  $^{12}$CH$_3$F}},
\newblock \bibinfo{journal}{Journal of Molecular Spectroscopy}
  \bibinfo{volume}{159} (\bibinfo{year}{1993}) \bibinfo{pages}{33--41}.
\bibitem[{{Nelson Jr.} et~al.(1967){Nelson Jr.}, Lide, and Maryott}]{67NeLiMa}
\bibinfo{author}{R.~D. {Nelson Jr.}}, \bibinfo{author}{D.~R. Lide},
  \bibinfo{author}{A.~A. Maryott}, \bibinfo{title}{Selected values of electric
  dipole moments for molecules in the gas phase}, \bibinfo{publisher}{U.S.
  Dept. of Commerce, National Bureau of Standards},
  \bibinfo{address}{Washington}, \bibinfo{year}{1967}.
\bibitem[{Cowan and Griffin(1976)}]{76CoGr}
\bibinfo{author}{R.~D. Cowan}, \bibinfo{author}{D.~C. Griffin},
\newblock \bibinfo{title}{{Approximate relativistic corrections to atomic
  radial wave functions}} \bibinfo{volume}{66} (\bibinfo{year}{1976})
  \bibinfo{pages}{1010--1014}.
\bibitem[{Balasubramanian(1997)}]{97Balasubr}
\bibinfo{author}{K.~Balasubramanian}, \bibinfo{title}{{Relativistic Effects in
  Chemistry}}, \bibinfo{publisher}{Wiley, New York}, \bibinfo{year}{1997}.
\bibitem[{Tarczay et~al.(2001)Tarczay, Cs\'asz\'ar, Klopper, and
  Quiney}]{01TaCsKlQu}
\bibinfo{author}{G.~Tarczay}, \bibinfo{author}{A.~G. Cs\'asz\'ar},
  \bibinfo{author}{W.~Klopper}, \bibinfo{author}{H.~M. Quiney},
\newblock \bibinfo{title}{{Anatomy of relativistic energy corrections in light
  molecular systems}},
\newblock \bibinfo{journal}{Mol. Phys.} \bibinfo{volume}{99}
  (\bibinfo{year}{2001}) \bibinfo{pages}{1769--1794}.
\bibitem[{Herzberg(1966)}]{66Herzberg}
\bibinfo{author}{G.~Herzberg}, \bibinfo{title}{Molecular Spectra and Molecular
  Structure: Electronic Spectra and Electronic Structure of Polyatomic
  Molecules}, \bibinfo{publisher}{D. van Nostrand, New York},
  \bibinfo{year}{1966}.
\bibitem[{Kandalam et~al.(2015)Kandalam, Kiran, Jena, Pietsch, and
  Gantef{\"o}r}]{kandalam2015superhalogens}
\bibinfo{author}{A.~K. Kandalam}, \bibinfo{author}{B.~Kiran},
  \bibinfo{author}{P.~Jena}, \bibinfo{author}{S.~Pietsch},
  \bibinfo{author}{G.~Gantef{\"o}r},
\newblock \bibinfo{title}{Superhalogens beget superhalogens: a case study of
  ({BO}$_2$)$_n$ oligomers},
\newblock \bibinfo{journal}{Phys. Chem. Chem. Phys.} \bibinfo{volume}{17}
  (\bibinfo{year}{2015}) \bibinfo{pages}{26589--26593}.
\bibitem[{Pritchard and Skinner(1955)}]{pritchard1955concept}
\bibinfo{author}{H.~Pritchard}, \bibinfo{author}{H.~Skinner},
\newblock \bibinfo{title}{The concept of electronegativity},
\newblock \bibinfo{journal}{Chem. Rev.} \bibinfo{volume}{55}
  (\bibinfo{year}{1955}) \bibinfo{pages}{745--786}.
\bibitem[{Bevington and Robinson(2003)}]{bevington2003data}
\bibinfo{author}{P.~R. Bevington}, \bibinfo{author}{D.~K. Robinson},
  \bibinfo{title}{Data Reduction and Error Analysis for the Physical Sciences,
  Third Edition}, \bibinfo{publisher}{McGraw-Hill}, \bibinfo{address}{New
  York}, \bibinfo{year}{2003}.
\bibitem[{Szab\'o et~al.(2020)Szab\'o, G\'oger, Charry, Karimpour, Fedorov, and
  Tkatchenko}]{20SzGoChKa}
\bibinfo{author}{P.~Szab\'o}, \bibinfo{author}{S.~G\'oger},
  \bibinfo{author}{J.~Charry}, \bibinfo{author}{M.~R. Karimpour},
  \bibinfo{author}{D.~V. Fedorov}, \bibinfo{author}{A.~Tkatchenko},
  \bibinfo{title}{Four-dimensional scaling of dipole polarizability in quantum
  systems}. \href{http://arxiv.org/abs/2010.11809}{{\tt arXiv:2010.11809}}.
\bibitem[{Brabec and Krausz(2000)}]{00BrKr}
\bibinfo{author}{T.~Brabec}, \bibinfo{author}{F.~Krausz},
\newblock \bibinfo{title}{Intense few-cycle laser fields: Frontiers of
  nonlinear optics},
\newblock \bibinfo{journal}{Rev. Mod. Phys.} \bibinfo{volume}{72}
  (\bibinfo{year}{2000}) \bibinfo{pages}{545--591}.

\end{thebibliography}

\end{document}